\documentclass[10pt,aps,prd,preprintnumbers,groupedaddress,showpacs,showkeys]{revtex4-2}
\usepackage{amsmath,amssymb,graphicx,epsfig,color,xcolor,ulem,cancel,datetime,dcolumn,bm}
\def\comment#1{}

\def\nn{\nonumber}

\def\dsf{\mathrm{dS_{4}}}
\def\tot{\mathrm{tot}}
\def\gr{\mathrm{gr}}
\def\elm{\mathrm{em}}
\def\x{\mathbf{x}}
\def\k{\mathbf{k}}
\def\xib{\bar{\xi}}
\def\vin{\mathrm{in}}
\begin{document}
%\preprint{arXiv}
\title{Induced energy-momentum tensor in de~Sitter scalar QED and its implication for induced current}
%\thanks{???}
\author{Omid~Gholizadeh~Meimanat}
%\email{o.gholizadeh22@gmail.com}
\author{Ehsan Bavarsad}
\email{bavarsad@kashanu.ac.ir}
\affiliation{Department of Physics, University of Kashan, 8731753153 Kashan, Iran}
%\date{\blue{\today,~\currenttime}}
\begin{abstract}
The aim of this research is to investigate the vacuum energy-momentum tensor of a quantized, massive, nonminimally coupled scalar field induced by a uniform electric field background in a four-dimensional de~Sitter spacetime ($\dsf$). We compute the expectation value of the energy-momentum tensor in the in-vacuum state and then regularize it using the adiabatic subtraction procedure. The correct trace anomaly of the induced energy-momentum tensor that confirmed our results is significant. The nonconservation equation for the induced energy-momentum tensor imposes the renormalization condition for the induced electric current of the scalar field. The findings of this research indicate that there are significant differences between the two induced currents which are regularized by this renormalization condition and the minimal subtraction condition.
\end{abstract}
%\pacs{???}
%\keywords{???}
\maketitle
%\tableofcontents
%%%%%%%%%%%%%%%%%%%%%%%%%%%%%%%%%%%%%%%%%%%%%%%%%%%%%%%%%%%%%%%%%%%%%%%%%%%%%%%%%%%%%%%%%%%%%%%%%%%%%%%%%%%%%%%%%%%%%%%%%%%%%%%%%%%%%%%%%%%%%%%%%%%%%%%%%%%%%%%
%%%%%%%%%%%%%%%%%%%%%%%%%%%%%%%%%%%%%%%%%%%%%%%%%%%%%%%%%%%%%%%%%%%%%%%%%%%%%%%%%%%%%%%%%%%%%%%%%%%%%%%%%%%%%%%%%%%%%%%%%%%%%%%%%%%%%%%%%%%%%%%%%%%%%%%%%%%%%%%
\section{\label{sec:intro}introduction}
In the past several decades quantum field theory in curved spacetime has played a major role in the study of the effects of gravity on quantum fields; for an pedagogical introduction see \cite{Book:Birrell,Book:Parker,Book:Wald}. A precise understanding of the quantum effects in curved spacetime was acquired by the late 1960s by Parker \cite{Parker:1968mv,Parker:1969au,Parker:1971pt} and followed by investigations of others. Indeed, these early investigations focused primarily on the physical consequences of particle creation in the cosmological spacetimes. It has been illustrated that a time-varying gravitational field creates elementary particles from the vacuum. Parker discovered that this particle creation process can be analyzed by using the Bogoliubov transformations method \cite{Parker:1969au}. Formulating a general framework of quantum field theory in curved spacetime involves nontrivial questions. The problem of particle concept is a deep one, and it is associated with one of the most fundamental difficulties of quantum field theory in curved spacetime, that there is no an unambiguous or unique vacuum state; a detailed discussion can be found in \cite{Book:Birrell,Book:Parker}. This ambiguity is reflected in the theory by the absence of an unambiguous or unique preferred mode solutions of the field equation, which is in turn a consequence of symmetry group of the spacetime. Since in a general nonstatic curved spacetime there is generically no any timelike Killing vector field, it is not possible to classify modes as positive-frequency or negative-frequency. Indeed, it is possible to construct a compleat set of modes, however the problem is that there are many of such sets and we will not have any criterion available to select a unique privileged choice of the modes. One of the key lessens learned from the development of this subject is that the notion of particle number does not generally have universal significant. Hence, the expectation value of the energy-momentum tensor in a suitable vacuum state is perhaps more physically relevant object to probe the structure of quantum fields in curved spacetime rather than particle number quantity \cite{Book:Birrell}. Part of reason is that the expectation value of the energy-momentum tensor in a fixed vacuum state transforms according to the usual tensor transformation law. In particular, if the vacuum expectation value of the energy-momentum tensor vanishes for an observer, it will vanish for all observers. On the contrary, the expectation of particle number is an essential observer-dependent quantity. The energy-momentum tensor is also important, because it is directly relevant to explore the consequences of the quantum field
dynamics for the geometry of the spacetime through the Einstein equation. Hence, it is interesting to investigate the expectation value of the energy-momentum tensor in a suitable vacuum state.
\par
Since the mid-seventies, significant advances have been made in the computation of the energy-momentum tensor of the quantum fields as well as its applications in the cosmological context. An essential feature of these computations is that they all involve the ultraviolet divergencies. In fact, such divergencies occur naturally in quantum field theory calculations. Various prescriptions have been developed for rendering the energy-momentum tensor finite. Among these prescriptions, Pauli-Villars regularization \cite{Bernard:1977pq,Vilenkin:1978wc}, dimensional regularization \cite{Brown:1976wc,
Candelas:1975du,Dowker:1975tf}, and zeta-function regularization \cite{Dowker:1975tf,Dowker:1976zf,Hawking:1976ja,Moretti:1997qn} were originally developed for use in Minkowski spacetime. Several sets of fictitious fields may be necessary to remove all of the divergences, making Pauli-Villars regularization rather complicated. In \cite{Hawking:1976ja}, it was pointed out that dimensional regularization procedure is ambiguous in curved spacetime due to the fact that in some classes of spacetimes, there is no natural way to generalize the dimensionality of the spacetime and the answer would be different in different extensions to $D$ dimensions. This method also suffers from limitation that one needs, in principle, to solve the field equations exactly. Another often used regularization procedure is point-splitting technique \cite{Christensen:1976vb,Adler:1976jx,Christensen:1978yd,Wald:1978pj} which is intrinsically designed to be used in the position-space representation of the composite operators. This regularization is implemented by placing the two quantum fields at distant points separated by an infinitesimal distance in a nonnull direction and then the divergencies that arise in the coincidence limit are absorbed into the renormalized parameters. Although point-splitting technique is the most applicable regularization scheme, it involves considerable technical complication. An alternative regularization prescription is adiabatic subtraction \cite{Parker:1974qw,Fulling:1974zr,Fulling:1974pu,Birrell:1978ap,
Hu:1978ap} which was originally invented by Parker \cite{Parker:1969au} to obtain the average density of the scalar particles created in a spatially flat Friedmann-Lemaitre-Robertson-Walker (FLRW) universe. This approach is only applicable in spacetimes with slowly varying curvature, and in fact an adiabatic expansion is an expansion in number derivatives of the spacetime metric. Hence, it is especially useful for studies that involve numerical techniques and problems of cosmological interest \cite{Birrell:1978ap}. In order to arrive a sensible semiclassical backreaction theory of quantum fields in curved spacetime, Wald has propounded \cite{Wald:1977up} a minimal set of properties that the renormalized energy-momentum tensor should satisfy. These properties which are often called the Wald axioms, in the weaker set of axioms, can be expressed as follows \cite{Book:Birrell,Book:Wald}: (1) The expectation value of the energy-momentum tensor in any state at a point $p$ in spacetime is covariant under general coordinate transformations (diffeomorphisms) and is independent of spacetime geometry at any point $q\neq p$. (2) The off-diagonal matrix element of energy-momentum tensor between any orthogonal pairs of states is finite and unambiguous. (3) For all states, the expectation value of the energy-momentum tensor is covariantly conserved. (4) The expectation value of the energy-momentum tensor vanishes in the relevant vacuum state in the Minkowski spacetime. It was shown in \cite{Wald:1978pj} that the fifth axiom of Ref.~\cite{Wald:1977up} cannot be satisfied. Wald then proved a remarkable result, i.e., the uniqueness theorem, which states that if the expectation value of the energy-momentum tensor satisfies the Wald axioms (1)-(3), then it is unique up to the addition of a local conserved tensor \cite{Wald:1977up}. Hence, the final results for the renormalized energy-momentum tensor do not depend on the employed regularization method \cite{Book:Birrell,Book:Wald}. Using these regularization techniques, the regularized and renormalized energy-momentum tensor of a neutral scalar field has been widely investigated in FLRW universes \cite{Parker:1974qw,Fulling:1974zr,Fulling:1974pu,Birrell:1978ap,Bunch:1978aq,Bunch:1977sq,Bunch:1978yw,
Davies:1977ze,Bunch:1978gb,Bunch:1980vc,Anderson:1987yt,Habib:1999cs,Zhang:2019gtg} and its physical implications for cosmological issues have been explored \cite{Markkanen:2017rvi,SolaPeracaula:2022hpd,Moreno-Pulido:2022phq}.
\par
It is thought that the very early universe can be approximately described by de~Sitter spacetime (dS) \cite{Book:Parker}, which motivates us to study the quantum field theory in this spacetime. Gibbons and Hawking \cite{Gibbons:1977mu} originally discovered that an inertial observer with a particle detector at rest perceives the de~Sitter invariant vacuum state as a bath of thermal radiation which apparently comes from the cosmological event horizon. In this context, a relatively simple model problem is that of a neutral scalar field with no self-interactions whose regularized and renormalized energy-momentum tensor has been analyzed \cite{Dowker:1975tf,Bunch:1978yq,Mottola:1984ar,Anderson:2013ila,Anderson:2013zia,Markkanen:2016aes,Anderson:2000wx,
Zhang:2019urk,Ye:2022tgs}. In \cite{Mottola:1984ar,Anderson:2013ila,Anderson:2013zia}, it was subsequently shown that the vacuum state of the quantum field in dS is unstable to particle creation. Furthermore, the study of semiclassical backreaction effect of the quantum corrections onto the Hubble rate in \cite{Markkanen:2016aes} indicated that these contributions can potentially result in a superacceleration phase, i.e., a phase when the Hubble rate increases as the dS expands. The spontaneous creation of pairs of charged particles from the vacuum by a strong electric field background in Minkowski spacetime is a well-known nonperturbative feature of quantum field theory \cite{Schwinger:1951nm,Heisenberg:1936eu,Sauter:1931zz}. Since the clearest version of this effect was worked out by Schwinger, it is named the Schwinger effect; reviews of this subject can be found in, e.g., \cite{Gelis:2015kya,Dunne:2004nc}. The phenomenon of particle creation has revealed the close analogy between quantum field theory in dS and a constant, uniform electric field in Minkowski spacetime \cite{Anderson:2013ila,Anderson:2013zia,Antoniadis:2006wq,Martin:2007bw}. This analogy motivates us to explore the combined implications of the dS Gibbons-Hawking effect and the Schwinger effect. Aside from this analogy, there are arguments and evidences that require the existence of strong electromagnetic fields in the early universe \cite{Maleknejad:2012fw,Durrer:2013pga}. This logical possibility provides a further reason for studying quantum field theory in the presence of an electric field in dS.
\par
There has been numerous studies to investigate the Schwinger pair creation by a uniform electric field background with the constant energy density in a dS. Seminal contributions have been made by \cite{Garriga:1993fh,Garriga:1994bm}. The Bogoliubov transformation method has been used to analyze the Schwinger effect for the charged scalar field in two- \cite{Garriga:1993fh,Garriga:1994bm,Kim:2008xv,Cai:2014qba,Hamil:2018rvu,Frob:2014zka}, four- \cite{Kobayashi:2014zza}, and arbitrary-dimensional \cite{Bavarsad:2016cxh} de~Sitter spacetimes. The Bogoliubov transformation method has been used to investigate the Schwinger effect for the Dirac field in dS \cite{Villalba:1995za,Haouat:2012ik,Stahl:2015gaa,Haouat:2015uaa,Stahl:2015cra}, in a manner similar to that used for the corresponding charged scalar field. In this method of analysis, which bring its own insights, the expectation value for the
number of particles is related to the Bogoliubov coefficients which, in turn, can be determined by specifying the in-vacuum and out-vacuum states. A reasonable definition of the out-vacuum state requires that the parameters mass of the quantum field and the electric field strength to satisfy the semiclassical condition. For a created semiclassical particle, this condition implies that either or both the mass and the magnitude of its electric potential energy acroses one Hubble radius must be much greater than the energy scale determined by the spacetime curvature \cite{Frob:2014zka,Kobayashi:2014zza}. An immediate consequence of the semiclassical condition is that the entire physical ranges of the parameters mass and electric field strength cannot be probed by this method. A useful alternative approach for studying the Schwinger effect in dS was introduced in Ref.~\cite{Frob:2014zka}. In this approach the expectation values of the objects such as the electric current and the energy-momentum tensor of the quantum field in the in-vacuum state are investigated. The choose of the in-state as the vacuum is justified form various viewpoints; see \cite{Garriga:1993fh}. Regarding the regularity properties, it is a Hadamard and adiabatic state \cite{Garriga:1994bm,Frob:2014zka}. The regularized expectation value of the electric current (also called induced current) in the in-vacuum state of the charged scalar field coupled to a constant, uniform electric field background has been evaluated in two- \cite{Frob:2014zka}, three- \cite{Bavarsad:2016cxh}, and four-dimensional \cite{Kobayashi:2014zza,Hayashinaka:2016dnt} de~Sitter spacetimes. In those analyses the behavior of the induced current can be probed in the infrared regime for which the quantum field mass is smaller than the magnitude of the electric potential energy acroses one Hubble radius which in turn is smaller than the energy scale determined by the spacetime curvature. The authors found that, although the induced current has been computed using different regularization method, in the infrared regime the induced current increases with deceasing electric field strength \cite{Frob:2014zka,Kobayashi:2014zza,Bavarsad:2016cxh,Hayashinaka:2016dnt}. This peculiar behavior was first observed in \cite{Frob:2014zka} and is called the infrared hyperconductivity. In \cite{Geng:2017zad}, the authors gave an alternative derivation of the infrared hyperconductivity phenomenon in $\dsf$ using the uniform asymptotic approximation method. For a charged scalar field coupled to a uniform electromagnetic field background with a constant energy density electric field parallel to a conserved flux magnetic field in $\dsf$, the Schwinger effect using Bogoliubov transformation method \cite{Bavarsad:2017oyv,Bavarsad:2018lvn,Moradi:2009zz} and the induced current \cite{Bavarsad:2017oyv,
Bavarsad:2018lvn} in the in-vacuum state have been investigated; and a period of infrared hyperconductivity was found. The in-vacuum induced current of the Dirac field coupled to a constant, uniform electric field background in two- \cite{Stahl:2015gaa} and four-dimensional \cite{Hayashinaka:2016qqn} de~Sitter spacetimes has been analyzed in a manner similar to that used for the corresponding scalar field. It was subsequently found that, in contrast to the case of corresponding scalar field, the infrared hyperconductivity phenomenon does not occur in the induced current of the Dirac field. Another peculiar behavior of the induced current occurs when the direction of the induced current is opposite the direction of applied electric field background. This phenomenon which is called the negative current, has been reported for both the scalar field \cite{Kobayashi:2014zza,Hayashinaka:2016dnt} with essentially small mass and the Dirac field \cite{Hayashinaka:2016qqn} with any mass in the four-dimensional de~Sitter spacetime. For the scalar field case the negative current occurs in a finite interval of the electric field strength, and for the Dirac field case it occurs below a certain value of the electric field strength which depends on the mass. In Refs.~\cite{Banyeres:2018aax,Hayashinaka:2018amz}, some notable attempts have been made to explain and remedy these peculiarities of the induced current. Beside the induced current, the regularized expectation value of the energy-momentum tensor (also called the induced energy-momentum tensor) in the in-vacuum state of the charged scalar \cite{AkbariAhmadmahmoudi:2021tpj,Bavarsad:2018mor,Bavarsad:2019jlg} and Dirac \cite{Botshekananfard:2019zak} quantum fields coupled to a constant, uniform electric field background has been analyzed in different dimensions of dS. In two-dimensional dS, the induced energy-momentum tensor of the charged scalar field has been analyzed in \cite{AkbariAhmadmahmoudi:2021tpj}, and subsequently the nonperturbative, regularized, one-loop effective Lagrangian of scalar QED has been constructed; the induced energy-momentum tensor of the corresponding Dirac field has been derived in \cite{Botshekananfard:2019zak}, and then applied to study of the gravitational backreaction effect. The trace of the induced energy-momentum tensor of the charged, massive scalar field conformally coupled to the Ricci scalar curvature of three- \cite{Bavarsad:2018mor} and four-dimensional \cite{Bavarsad:2019jlg} de~Sitter spacetimes has been computed; and to examine the evolution of the Hubble constant, the induced energy-momentum tensor has been obtained from the trace, along with the assumption that the created pairs act like a perfect fluid with a vacuum equation of state. By applying the Bogoliubov transformation method, the energy-momentum tensor of the Schwinger scalar pairs created by a constant, uniform electric field background in an arbitrary dimensional dS has been computed under the two limiting conditions, i.e., the heavy scalar field \cite{Bavarsad:2016cxh} and the strong electric field \cite{Bavarsad:2017wbe}. In both these cases, it was found that the Hubble constant decays as a consequence of the Schwinger pair creation. Before closing the present section, it is worthwhile to mention that the Schwinger effect has been studied in FLRW spacetimes \cite{Haouat:2012dr} and in the framework of cosmological models \cite{Geng:2017zad,Shakeri:2019mnt,Stahl:2018idd,Giovannini:2018qbq,Sharma:2017ivh,
Domcke:2019qmm,Tangarife:2017rgl,Kitamoto:2018htg,Chua:2018dqh,Gorbar:2019fpj,Sobol:2019xls,Sobol:2018djj}. The role of strong electromagnetic fields in astrophysics and cosmology was reviewed in \cite{Kim:2019joy}. The objectives of this article are to investigate the induced energy-momentum tensor of the charged scalar field coupled to a uniform electric field background in $\dsf$, and to analyze its nonconservation equation. The importance and originality of this study are that it calculates the induced energy-momentum tensor and explores a relation between the induced energy-momentum tensor and the induced current which leads to new insights into the regularization and behavior of the induced current.
\par
The remaining part of the article proceeds as follows: In Sec.~\ref{sec:basic}, we define and construct the basic elements of the formalism that will be necessary in our subsequent discussions. We then carry out explicit computation of the induced energy-momentum tensor in Sec.~\ref{sec:const}. The properties of the induced energy-momentum tensor are explored in detail in Sec.~\ref{sec:prob}. In Sec.~\ref{sec:curre}, we analyze the nonconservation equation of the induced energy-momentum tensor, this investigation yields a relation between the induced energy-momentum tensor and the induced current. Then the properties of the resulting induced current are discussed in detail. In Sec.~\ref{sec:concl}, we present the findings of the research. In the Appendix we present further supplementary data associated with the calculation of the expectation values of the components of the energy-momentum tensor.
%%%%%%%%%%%%%%%%%%%%%%%%%%%%%%%%%%%%%%%%%%%%%%%%%%%%%%%%%%%%%%%%%%%%%%%%%%%%%%%%%%%%%%%%%%%%%%%%%%%%%%%%%%%%%%%%%%%%%%%%%%%%%%%%%%%%%%%%%%%%%%%%%%%%%%%%%%%%%%%
%%%%%%%%%%%%%%%%%%%%%%%%%%%%%%%%%%%%%%%%%%%%%%%%%%%%%%%%%%%%%%%%%%%%%%%%%%%%%%%%%%%%%%%%%%%%%%%%%%%%%%%%%%%%%%%%%%%%%%%%%%%%%%%%%%%%%%%%%%%%%%%%%%%%%%%%%%%%%%%
\section{\label{sec:basic}basic definitions and constructions}
In this section we shall define and construct the basic elements of the formalism that will be necessary in our subsequent discussions.
%%%%%%%%%%%%%%%%%%%%%%%%%%%%%%%%%%%%%%%%%%%%%%%%%%%%%%%%%%%%%%%%%%%%%%%%%%%%%%%%%%%%%%%%%%%%%%%%%%%%%%%%%%%%%%%%%%%%%%%%%%%%%%%%%%%%%%%%%%%%%%%%%%%%%%%%%%%%%%%
%%%%%%%%%%%%%%%%%%%%%%%%%%%%%%%%%%%%%%%%%%%%%%%%%%%%%%%%%%%%%%%%%%%%%%%%%%%%%%%%%%%%%%%%%%%%%%%%%%%%%%%%%%%%%%%%%%%%%%%%%%%%%%%%%%%%%%%%%%%%%%%%%%%%%%%%%%%%%%%
\subsection{\label{sec:model}Specification of the model}
We have already mentioned that we consider a massive complex scalar field $\varphi(x)$, coupled to the electromagnetic vector potential $A_{\mu}(x)$,
which describes a uniform electric field background with a constant energy density in the conformal Poincar\'e patch of $\dsf$. To represent this region of $\dsf$ which is conformally related to a region of Minkowski spacetime, we choose the coordinates $x^{\mu}=(\tau,\mathbf{x})$ that the ranges of the conformal time $\tau$, and the comoving spatial coordinates $\mathbf{x}$ are given by
\begin{align}\label{ranges}
\tau &\in \big(-\infty,0\big), & \mathbf{x} &\in \mathbb{R}^{3}.
\end{align}
In terms of these coordinates the metric of the spacetime takes the form
\begin{equation}\label{metric}
g_{\mu\nu}dx^{\mu}dx^{\nu}=\Omega^{2}(\tau)\big(d\tau^{2}-d\mathbf{x}\cdot d\mathbf{x}\big),
\end{equation}
with the conformal scale factor
\begin{equation}\label{scale}
\Omega(\tau)=-\frac{1}{H\tau},
\end{equation}
where $H$ is the Hubble constant. The nonzero Christoffel symbols for the metric (\ref{metric}) are given by
\begin{align}\label{christoff}
\Gamma^{0}_{00}&=\frac{\dot{\Omega}}{\Omega}, & \Gamma^{0}_{ij}&= \frac{\dot{\Omega}}{\Omega}\delta_{ij}, &
\Gamma^{i}_{0j}&= \frac{\dot{\Omega}}{\Omega}\delta^{i}_{j},
\end{align}
where the roman indices $i,j$ denote only three spatial components and run from 1 to 3. We use the overdot to denote differentiation with respect to the conformal time $\tau$. From Eq.~(\ref{christoff}) the Ricci tensor and hence the Ricci scalar can be calculated
\begin{align}\label{ricci}
R_{\mu\nu}&=3H^{2}g_{\mu\nu}, & R&=12H^{2}.
\end{align}
We put a uniform electric field background with a constant energy density on the Poincar\'e patch represented in Eq.~(\ref{metric}). Without loss of generality, we choose our coordinates so that this electric field to point in the $x^{1}$ direction. Thus the nonzero components of the electromagnetic field tensor are
\begin{equation}\label{electric}
F_{01}=-F_{10}=\Omega^{2}(\tau)E,
\end{equation}
where $E$ is a constant. It is convenient to express this electromagnetic field tensor in terms of a vector potential in the Coulomb gauge \cite{Frob:2014zka,Kobayashi:2014zza,Bavarsad:2016cxh} as
\begin{equation}\label{gauge}
A_{\mu}(\tau)=-\frac{E}{H^{2}\tau}\delta^{1}_{\mu}.
\end{equation}
The compleat action $S_{\tot}$ of this theory can be represented as sum of a pure gravitational piece, an electromagnetic piece, and a scalar field piece
\begin{equation}\label{total}
S_{\tot}=S_{\gr}+S_{\elm}+S.
\end{equation}
Here $S_{\gr}$ is the Einstein-Hilbert action that only includes the gravitation piece of the compleat action, and is given by
\begin{equation}\label{hilbert}
S_{\gr}=\frac{1}{16\pi G}\int d^{4}x\sqrt{-g}\Big(R-2\Lambda_{\mathrm{c}}\Big),
\end{equation}
where $G$ is Newton's gravitational constant, $g$ is the determinant of the metric, $R$ is the Ricci scalar of the spacetime, and $\Lambda_{\mathrm{c}}$
is the cosmological constant. The electromagnetic piece of the compleat action is expressed in terms of the electromagnetic field tensor as
\begin{equation}\label{maxwell}
S_{\elm}=-\frac{1}{4}\int d^{4}x\sqrt{-g}F_{\mu\nu}F^{\mu\nu}.
\end{equation}
The dynamics of the complex scalar field $\varphi(x)$ of mass $m$ coupled to the electromagnetic vector potential (\ref{gauge}) with coupling $e$ is governed by the last piece of the compleat action which can be written as
\begin{equation}\label{sqed}
S=\int d^{4}x\sqrt{-g}\Big[g^{\mu\nu}\big(\partial_{\mu}+ieA_{\mu}\big)\varphi\big(\partial_{\nu}-ieA_{\nu}\big)\varphi^{\ast}
-(m^{2}+\xi R)\varphi\varphi^{\ast} \Big],
\end{equation}
where $\xi$ is a dimensionless nonminimal coupling constant which describes the strength of the coupling $\varphi$ to the Ricci scalar (\ref{ricci}).
The Euler-Lagrange equations of motion give the Klein-Gordon equation for the scalar field
\begin{equation}\label{eq:kg}
\frac{1}{\sqrt{-g}}\partial_{\mu}\big(\sqrt{-g}g^{\mu\nu}\partial_{\nu}\varphi\big)+2ieg^{\mu\nu}A_{\mu}\partial_{\nu}\varphi
-e^{2}g^{\mu\nu}A_{\mu}A_{\nu}\varphi+(m^{2}+\xi R)\varphi=0,
\end{equation}
and the Maxwell equation for the electromagnetic field
\begin{equation}\label{eq:maxwe}
\nabla_{\nu}F^{\nu\mu}=j^{\mu},
\end{equation}
where $\nabla$ denotes the covariant derivative operator, and $j^{\mu}$ is the electric current of the scalar field caused by the electric field background (\ref{electric}) which is defined by
\begin{equation}\label{def:j}
j^{\mu}(x)=ieg^{\mu\nu}\Big[\big(\partial_{\nu}\varphi+ieA_{\nu}\varphi\big)\varphi^{\ast}
-\varphi\big(\partial_{\nu}\varphi^{\ast}-ieA_{\nu}\varphi^{\ast}\big)\Big].
\end{equation}
It is straightforward to verify that $\nabla_{\mu}j^{\mu}=0$, i.e., the electric current is conserved.
%%%%%%%%%%%%%%%%%%%%%%%%%%%%%%%%%%%%%%%%%%%%%%%%%%%%%%%%%%%%%%%%%%%%%%%%%%%%%%%%%%%%%%%%%%%%%%%%%%%%%%%%%%%%%%%%%%%%%%%%%%%%%%%%%%%%%%%%%%%%%%%%%%%%%%%%%%%%%%%
%%%%%%%%%%%%%%%%%%%%%%%%%%%%%%%%%%%%%%%%%%%%%%%%%%%%%%%%%%%%%%%%%%%%%%%%%%%%%%%%%%%%%%%%%%%%%%%%%%%%%%%%%%%%%%%%%%%%%%%%%%%%%%%%%%%%%%%%%%%%%%%%%%%%%%%%%%%%%%%
\subsection{\label{sec:preli}Preliminary definition of the energy-momentum tensor}
The classical Einstein equation can be derived from the compleat action (\ref{total}) by demanding the invariance of $S_{\tot}$ under infinitesimal variation of the metric $\delta g_{\mu\nu}$, or equivalently, infinitesimal variation of the inverse metric $\delta g^{\mu\nu}$.
This condition requires that
\begin{equation}\label{delta:tot}
\frac{2}{\sqrt{-g}}\frac{\delta S_{\tot}}{\delta g^{\mu\nu}}=\frac{2}{\sqrt{-g}}\bigg(\frac{\delta S_{\gr}}{\delta g^{\mu\nu}}
+\frac{\delta S_{\elm}}{\delta g^{\mu\nu}}+\frac{\delta S}{\delta g^{\mu\nu}}\bigg)=0.
\end{equation}
The Variation of expression (\ref{hilbert}) with respect to $\delta g^{\mu\nu}$ leads to
\begin{equation}\label{delta:gr}
\frac{2}{\sqrt{-g}}\frac{\delta S_{\gr}}{\delta g^{\mu\nu}}
=\frac{1}{8\pi G}\bigg(R_{\mu\nu}-\frac{1}{2}Rg_{\mu\nu}+\Lambda_{\mathrm{c}} g_{\mu\nu}\bigg).
\end{equation}
The variations of the electromagnetic action $S_{\elm}$, and the scalar field action $S$, with respect to $\delta g^{\mu\nu}$ define the energy-momentum tensor of the electromagnetic field $T^{(\elm)}_{\mu\nu}$, and the energy-momentum tensor of the scalar field $T_{\mu\nu}$, respectively, as
\begin{eqnarray}
\frac{2}{\sqrt{-g}}\frac{\delta S_{\elm}}{\delta g^{\mu\nu}} &=& T^{(\elm)}_{\mu\nu}, \label{delta:em} \\
\frac{2}{\sqrt{-g}}\frac{\delta S}{\delta g^{\mu\nu}} &=& T_{\mu\nu}. \label{delta:sc}
\end{eqnarray}
Plugging expressions (\ref{maxwell}) and (\ref{sqed}) into Eqs.~(\ref{delta:em}) and (\ref{delta:sc}), respectively, and following the standard calculus
of variations procedure yields the energy-momentum tensor of the electromagnetic field
\begin{equation}\label{emt:em}
T^{(\elm)}_{\mu\nu}=\frac{1}{4}g_{\mu\nu}F_{\rho\sigma}F^{\rho\sigma}+g^{\rho\sigma}F_{\mu\rho}F_{\sigma\nu},
\end{equation}
and a preliminary expression for the energy-momentum tensor of the scalar field
\begin{eqnarray}\label{emt:sc}
T_{\mu\nu} &=& \Big[\big(4\xi-1\big)g^{\rho\sigma}\big(\partial_{\rho}+ieA_{\rho}\big)\varphi\big(\partial_{\sigma}-ieA_{\sigma}\big)\varphi^{\ast}
+\big(1-4\xi\big)m^{2}\varphi\varphi^{\ast}+\Big(\frac{1}{2}-4\xi\Big)\xi R\varphi\varphi^{\ast}\Big]g_{\mu\nu} \nn\\
&+&\big(1-2\xi\big)\Big(\partial_{\mu}\varphi\partial_{\nu}\varphi^{\ast}+\partial_{\nu}\varphi\partial_{\mu}\varphi^{\ast}\Big)
+ieA_{\mu}\Big(\varphi\partial_{\nu}\varphi^{\ast}-\partial_{\nu}\varphi\varphi^{\ast}\Big)
+ieA_{\nu}\Big(\varphi\partial_{\mu}\varphi^{\ast}-\partial_{\mu}\varphi\varphi^{\ast}\Big) \nn\\
&+&2e^{2}A_{\mu}A_{\nu}\varphi\varphi^{\ast}+2\xi\Gamma^{\rho}_{\mu\nu}\Big(\varphi\partial_{\rho}\varphi^{\ast}+\partial_{\rho}\varphi\varphi^{\ast}\Big)
-2\xi\Big(\partial_{\mu}\partial_{\nu}\varphi\varphi^{\ast}+\varphi\partial_{\mu}\partial_{\nu}\varphi^{\ast}\Big).
\end{eqnarray}
Plugging three expressions (\ref{delta:gr})-(\ref{delta:sc}) into Eq.~(\ref{delta:tot}) gives the Einstein equation
\begin{equation}\label{einstein}
R_{\mu\nu}-\frac{1}{2}Rg_{\mu\nu}+\Lambda_{\mathrm{c}}g_{\mu\nu}=-8\pi G\Big(T^{(\elm)}_{\mu\nu}+T_{\mu\nu}\Big).
\end{equation}
An important property of the Einstein equation is that both sides of Eq.~(\ref{einstein}) have identically vanishing covariant divergences, which implies
\begin{equation}\label{divergen}
\nabla_{\mu}T^{\mu\nu}=-\nabla_{\mu}T^{(\elm)\mu\nu}.
\end{equation}
Using the Klein-Gordon Eq.~(\ref{eq:kg}), it is straightforward to show that the covariant divergence of expression (\ref{emt:sc}) is given by
\begin{equation}\label{diver:sc}
\nabla_{\mu}T^{\mu\nu}=-j_{\mu}F^{\mu\nu},
\end{equation}
Apparently, the energy-momentum tensor of the scalar field is not covariantly conserved in the presence of the electromagnetic field, as the consequence
of the electromagnetic interactions. We show that the nonconservation of $T_{\mu\nu}$ is compatible with the nonconservation of $T^{(\elm)}_{\mu\nu}$ so that the relation (\ref{divergen}) is satisfied, and hence the total energy momentum tensor is covariantly conserved. By taking the covariant divergence of the expression (\ref{emt:em}) and using the Maxwell equation (\ref{eq:maxwe}), it is seen that
\begin{equation}\label{diver:em}
\nabla_{\mu}T^{(\elm)\mu\nu}=j_{\mu}F^{\mu\nu}.
\end{equation}
As a result of Eqs.~(\ref{diver:sc}) and (\ref{diver:em}), it is evident that the relation (\ref{divergen}) is satisfied. Hence, the total energy-momentum tensor $T_{\mu\nu}+T^{(\elm)}_{\mu\nu}$ is manifestly conserved.
\par
It will be important to state that we will treat the classical gravitational field (\ref{metric}) and the classical electromagnetic field (\ref{gauge}) as fixed field configurations which they are unaffected by the dynamics of the quantum complex scalar field $\varphi(x)$ in response to these backgrounds. In fact, the Einstein equation (\ref{einstein}) describes the backreaction effects on the gravitational field, and the Maxwell equation (\ref{eq:maxwe}) and Eq.~(\ref{diver:em}) describe the backreaction effects on the electromagnetic field. Indeed, due to the conceptual importance of Eq.~(\ref{diver:sc}) in
our subsequent discussions, we have presented Eqs.~(\ref{eq:maxwe}), (\ref{einstein}), and (\ref{diver:em}) to provide a fairly detailed discussion of its derivation. It is then clear that we do not discuss these equations further in this article.
%%%%%%%%%%%%%%%%%%%%%%%%%%%%%%%%%%%%%%%%%%%%%%%%%%%%%%%%%%%%%%%%%%%%%%%%%%%%%%%%%%%%%%%%%%%%%%%%%%%%%%%%%%%%%%%%%%%%%%%%%%%%%%%%%%%%%%%%%%%%%%%%%%%%%%%%%%%%%%%
%%%%%%%%%%%%%%%%%%%%%%%%%%%%%%%%%%%%%%%%%%%%%%%%%%%%%%%%%%%%%%%%%%%%%%%%%%%%%%%%%%%%%%%%%%%%%%%%%%%%%%%%%%%%%%%%%%%%%%%%%%%%%%%%%%%%%%%%%%%%%%%%%%%%%%%%%%%%%%%
\subsection{\label{sec:quant}Quantizing the complex scalar field}
Quantizing the complex scalar field $\varphi(x)$, in the classical gravitational (\ref{metric}) and electromagnetic (\ref{gauge}) field backgrounds is
completely straightforward and follows exactly the same route as that of a complex scalar field in Minkowski spacetime. We will solve the Kline-Gordon
Eq.~(\ref{eq:kg}), and then adopting canonical quantization method, we can define the in-vacuum state to obtain the expectation value of the
energy-momentum tensor operator.
\par
Taking account of the fact that the gravitational (\ref{metric}) and electromagnetic (\ref{gauge}) backgrounds are invariant under spatial translations,
it is convenient to write the mode solution of Eq.~(\ref{eq:kg}) as
\begin{equation}\label{ansatz}
U_{\k}(x)=\Omega^{-1}(\tau)e^{i\k\cdot\x}f(\tau),
\end{equation}
where $\k$ is the comoving momentum. Plugging Eqs.~(\ref{metric}), (\ref{gauge}), and (\ref{ansatz}) into Eq.~(\ref{eq:kg}), we can put the differential equation in a standard form by the change of variable $z=2ik\tau$ so that $k=|\k|$. Then it becomes
\begin{equation}\label{whitaker}
\frac{d^{2}f}{dz^{2}}+\bigg(-\frac{1}{4}+\frac{\kappa}{z}+\frac{1/4-\gamma^{2}}{z^{2}}\bigg)f(z)=0,
\end{equation}
where the dimensionless parameters are defined through
\begin{align}\label{paramete}
\mu&=\frac{m}{H}, & \lambda&=-\frac{eE}{H^{2}}, & \xib&=\xi-\frac{1}{6}, \nn \\
r&=\frac{k_{x}}{k}, & \kappa&=-i\lambda r, & \gamma&=\sqrt{\frac{1}{4}-\lambda^{2}-\mu^{2}-12\bar{\xi}}.
\end{align}
Two values of $\xi$ are of particular interest that are the minimally coupled case $\xi=0$, and conformally coupled case $\xi=1/6$ which implies $\xib=0$. The variable $k_{x}$ which appears in the definition of the parameter $r$, denotes the component of the comoving momentum $\k$ along the electric field background. Equation (\ref{whitaker}) is the Whittaker equation, and the solutions are called Whittaker functions; see, e.g., \cite{Book:NIST}. We define the in-vacuum state $|\vin\rangle$ so that in the remote past ($\tau\rightarrow-\infty$) where the spacetime is asymptotically Minkowskian, an inertial observer there would identify this state with a physical vacuum. This vacuum state may be represented by a mode solution of the Whittaker Eq.~(\ref{whitaker}) which behaves like a mode function in Minkowski spacetime in the limit of $\tau\rightarrow-\infty$. Hence, the solution of Eq.~(\ref{whitaker}) with the desired asymptotic form in the limit of $|z|\rightarrow\infty$ which can be represented in terms of a Mellin-Barnes integral \cite{Book:NIST} is
\begin{equation}\label{mellin}
W_{\kappa,\gamma}(z)=\frac{e^{-\frac{z}{2}}}{\Gamma\big(\frac{1}{2}+\gamma-\kappa\big)\Gamma\big(\frac{1}{2}-\gamma-\kappa\big)}
\int_{-i\infty}^{+i\infty}\frac{ds}{2\pi i}\Gamma\big(\frac{1}{2}+\gamma+s\big)\Gamma\big(\frac{1}{2}-\gamma+s\big)\Gamma\big(-\kappa-s\big)z^{-s},
\end{equation}
with the condition that the phase of the variable $z$ and the values of the parameters $\gamma$ and $\kappa$ must satisfy the following inequalities
\begin{align}\label{inequalit}
\big|\mathrm{ph}(z)\big|&<\frac{3\pi}{2}, & \frac{1}{2}\pm\gamma-\kappa&\neq0,-1,-2,\cdots.
\end{align}
In expression (\ref{mellin}), the factors denoted by $\Gamma$ are the gamma functions. The contour of the Mellin-Barnes integral (\ref{mellin}) consists of a straight vertical line from minus infinity to infinity, parallel to imaginary axis in the complex plane, and of a semicircle at infinity with indentations if necessary to avoid poles of the integrand in a way that separates the poles of $\Gamma\big(\frac{1}{2}+\gamma+s\big)$ and $\Gamma\big(\frac{1}{2}-\gamma+s\big)$ from the poles of $\Gamma\big(-\kappa-s\big)$. The normalized mode functions which behave like the positive frequency
Minkowski mode functions in the remote past are given by \cite{Kobayashi:2014zza,Bavarsad:2016cxh},
\begin{equation}\label{umode}
U_{\k}(x)=\frac{1}{\sqrt{2k}}e^{\frac{i\pi\kappa}{2}}\Omega^{-1}(\tau)e^{i\k\cdot\x}W_{\kappa,\gamma}\big(2ik\tau\big).
\end{equation}
Besides, the normalized mode functions which behave like the negative frequency Minkowski mode functions in the remote past are found to be \cite{Kobayashi:2014zza,Bavarsad:2016cxh},
\begin{equation}\label{vmode}
V_{\k}(x)=\frac{1}{\sqrt{2k}}e^{-\frac{i\pi\kappa}{2}}\Omega^{-1}(\tau)e^{-i\k\cdot\x}W_{\kappa,\gamma}\big(-2ik\tau\big).
\end{equation}
We have conventionally normalized the mode functions (\ref{umode}) and (\ref{vmode}) such that their Wronskian to be
\begin{equation}\label{wronskia}
U_{\k}\dot{U}_{\k}^{\ast}-U_{\k}^{\ast}\dot{U}_{\k}=V_{\k}^{\ast}\dot{V}_{\k}-V_{\k}\dot{V}_{\k}^{\ast}=i\Omega^{-2}(\tau).
\end{equation}
These mode functions will be orthonormal with respect to the conserved scalar product integrated over the constant $\tau$ hypersurface \cite{Bavarsad:2016cxh}. The usual canonical quantization will proceed by introducing the creation $a_{\k}^{\dag}$, and annihilation $a_{\k}$ operators for each of mode functions $U_{\k}$, and similarly the creation $b_{\k}^{\dag}$, and annihilation $b_{\k}$ operators for each of mode functions $V_{\k}$. The creation and annihilation operators satisfy the commutation rules
\begin{equation}\label{commutat}
\Big[a_{\k},a_{\k'}^{\dag}\Big]=\Big[b_{\k},b_{\k'}^{\dag}\Big]=(2\pi)^{3}\delta\big(\k-\k'\big),
\end{equation}
with all other commutators vanishing. Then, the complex scalar field operator can be expanded in terms of the creation and annihilation operators in the standard manner as
\begin{equation}\label{operator}
\varphi(x)=\int\frac{d^{3}k}{(2\pi)^{3}}\Big[a_{\k}U_{\k}(x)+b_{\k}^{\dag}V_{\k}(x)\Big].
\end{equation}
The in-vacuum state $|\vin\rangle$ is characterized by the fact that it is annihilated by each of $a_{\k}$ and $b_{\k}$ operators
\begin{align}\label{vin}
a_{\k}\big|\vin\big\rangle=b_{\k}\big|\vin\big\rangle=0, && \forall \k.
\end{align}
%%%%%%%%%%%%%%%%%%%%%%%%%%%%%%%%%%%%%%%%%%%%%%%%%%%%%%%%%%%%%%%%%%%%%%%%%%%%%%%%%%%%%%%%%%%%%%%%%%%%%%%%%%%%%%%%%%%%%%%%%%%%%%%%%%%%%%%%%%%%%%%%%%%%%%%%%%%%%%%
%%%%%%%%%%%%%%%%%%%%%%%%%%%%%%%%%%%%%%%%%%%%%%%%%%%%%%%%%%%%%%%%%%%%%%%%%%%%%%%%%%%%%%%%%%%%%%%%%%%%%%%%%%%%%%%%%%%%%%%%%%%%%%%%%%%%%%%%%%%%%%%%%%%%%%%%%%%%%%%
\section{\label{sec:const}construction of the induced energy-momentum tensor}
Having built a foundation for our discussion and presented the definition for the energy-momentum tensor of the scalar field in Eq.~(\ref{emt:sc}), we will proceed to present the regularized in-vacuum expectation value of the energy-momentum tensor of the scalar field.
%%%%%%%%%%%%%%%%%%%%%%%%%%%%%%%%%%%%%%%%%%%%%%%%%%%%%%%%%%%%%%%%%%%%%%%%%%%%%%%%%%%%%%%%%%%%%%%%%%%%%%%%%%%%%%%%%%%%%%%%%%%%%%%%%%%%%%%%%%%%%%%%%%%%%%%%%%%%%%%
%%%%%%%%%%%%%%%%%%%%%%%%%%%%%%%%%%%%%%%%%%%%%%%%%%%%%%%%%%%%%%%%%%%%%%%%%%%%%%%%%%%%%%%%%%%%%%%%%%%%%%%%%%%%%%%%%%%%%%%%%%%%%%%%%%%%%%%%%%%%%%%%%%%%%%%%%%%%%%%
\subsection{\label{sec:unreg}Unregularized expectation values in the in-vacuum state}
Integral representations for the in-vacuum expectation values of the components of the energy-momentum tensor can be obtained by substituting the mode  expansion (\ref{operator}) for the quantum scalar field $\varphi(x)$ into the definition (\ref{emt:sc}), and then calculating the expectation values in the in-vacuum state using relations (\ref{commutat}) and (\ref{vin}). By using Eq.~(\ref{eq:kg}) and some algebra, we have the following integral expressions in terms of the positive frequency mode function (\ref{umode}) for the expectation values of the components. The integral expression of the timelike component is given by
\begin{equation}\label{vint00}
\big\langle\vin\big|T_{00}\big|\vin\big\rangle = \int\frac{d^{3}k}{(2\pi)^{3}}\bigg[\dot{U}_{\k}\dot{U}_{\k}^{\ast}-6\xi\tau^{-1}\Big(U_{\k}\dot{U}_{\k}^{\ast}+\dot{U}_{\k}U_{\k}^{\ast}\Big)
+\tau^{-2}\Big(k^{2}\tau^{2}+2\lambda rk\tau+\lambda^{2}+\mu^{2}
+6\xi\Big)U_{\k}U_{\k}^{\ast}\bigg].
\end{equation}
For the diagonal spacelike components we get
\begin{eqnarray}\label{vint11}
\big\langle\vin\big|T_{11}\big|\vin\big\rangle &=& \int\frac{d^{3}k}{(2\pi)^{3}}\bigg\{\big(1-4\xi\big)\dot{U}_{\k}\dot{U}_{\k}^{\ast}
-2\xi\tau^{-1}\Big(U_{\k}\dot{U}_{\k}^{\ast}+\dot{U}_{\k}U_{\k}^{\ast}\Big)
+\tau^{-2}\Big[\big(4\xi-1+2r^{2}\big)k^{2}\tau^{2}+2\big(4\xi+1\big)\lambda rk\tau \nn\\
&+&\big(4\xi+1\big)\lambda^{2}+\big(4\xi-1\big)\mu^{2}+6\xi\big(8\xi-1\big)\Big]U_{\k}U_{\k}^{\ast}\bigg\},
\end{eqnarray}
and
\begin{eqnarray}\label{vint22}
\big\langle\vin\big|T_{22}\big|\vin\big\rangle &=& \big\langle\vin\big|T_{33}\big|\vin\big\rangle
= \int\frac{d^{3}k}{(2\pi)^{3}}\bigg\{\big(1-4\xi\big)\dot{U}_{\k}\dot{U}_{\k}^{\ast}
-2\xi\tau^{-1}\Big(U_{\k}\dot{U}_{\k}^{\ast}+\dot{U}_{\k}U_{\k}^{\ast}\Big)
+\tau^{-2}\Big[\big(4\xi-1\big)k^{2}\tau^{2} \nn\\
&+&2k_{z}^{2}\tau^{2}+2\big(4\xi-1\big)\lambda rk\tau+\big(4\xi-1\big)\lambda^{2}+\big(4\xi-1\big)\mu^{2}
+6\xi\big(8\xi-1\big)\Big]U_{\k}U_{\k}^{\ast}\bigg\}.
\end{eqnarray}
The only nonvanishing in-vacuum expectation values of the off-diagonal components can be expressed as
\begin{equation}\label{vint01}
\big\langle\vin\big|T_{01}\big|\vin\big\rangle = \big\langle\vin\big|T_{10}\big|\vin\big\rangle
= i\tau^{-1}\int\frac{d^{3}k}{(2\pi)^{3}}\big(rk\tau+\lambda\big)\Big(U_{\k}\dot{U}_{\k}^{\ast}-\dot{U}_{\k}U_{\k}^{\ast}\Big).
\end{equation}
Using the asymptotic expansion of the Whittaker function (\ref{mellin}) for large values of its argument \cite{Book:NIST}, it can be shown that the mode function (\ref{umode}) is proportional to $k^{-\frac{1}{2}-i\lambda r}$ in the limit of $k\rightarrow\infty$. Then, inspection of integrals in
Eqs.~(\ref{vint00})-(\ref{vint01}) shows that these expressions are ultraviolet divergent. Thus, we first regulate them by cutting them of at a large momentum $K$. Further description of the calculation of the expressions (\ref{vint00})-(\ref{vint22}) is available in the Appendix. We find the final expression for the unregularized in-vacuum expectation value of the timelike component
\begin{flalign}\label{vinemt00}
&\big\langle\vin\big|T_{00}\big|\vin\big\rangle = \Omega^{2}(\tau)\frac{H^{4}}{8\pi^{2}}\bigg\{
\Lambda^{4}+\Big(\mu^{2}-6\xib+\frac{2\lambda^{2}}{3}\Big)\Lambda^{2}
-\Big(\frac{\mu^{4}}{2}+6\xib\mu^{2}-\frac{\lambda^{2}}{6}\Big)\log\big(2\Lambda\big)+54\xib^{2}
-\frac{2\lambda^{4}}{15}+\frac{\mu^{2}}{4}+6\xib\mu^{2}+\frac{\mu^{4}}{8} \nn\\
&-\frac{19\lambda^{2}}{72}-2\lambda^{2}\xib-\frac{\lambda^{2}\mu^{2}}{2}
+\frac{\gamma}{24\pi}\bigg(\frac{45}{\pi^{2}}-6-96\xib+4\lambda^{2}-26\mu^{2}\bigg)
\frac{\cosh\big(2\pi\lambda\big)}{\sin\big(2\pi\gamma\big)}
-\frac{\gamma}{48\pi^{2}\lambda}\bigg(\frac{45}{\pi^{2}}-6-96\xib+64\lambda^{2}-26\mu^{2}\bigg) \nn\\
&\times\frac{\sinh\big(2\pi\lambda\big)}{\sin\big(2\pi\gamma\big)}
+i\csc\big(2\pi\gamma\big)\int_{-1}^{+1}C_{0r}\bigg[\Big(e^{2\pi\lambda r}+e^{2i\pi\gamma}\Big)
\psi\Big(\frac{1}{2}+\gamma+i\lambda r\Big)
-\Big(e^{2\pi\lambda r}+e^{-2i\pi\gamma}\Big)\psi\Big(\frac{1}{2}-\gamma+i\lambda r\Big)\bigg]dr \nn\\
&+\frac{i\pi}{12}\Big(3\mu^{4}+36\xib\mu^{2}-\lambda^{2}\Big)
\bigg\},&&
\end{flalign}
where the coefficient $C_{0r}$ is given by
\begin{equation}\label{c0}
C_{0r}=-\frac{5}{8}\lambda^{4}r^{4}+\frac{1}{8}\Big(6\mu^{2}+6\lambda^{2}+36\xib+1\Big)\lambda^{2}r^{2}
-\frac{1}{8}\big(\mu^{2}+\lambda^{2}\big)\big(\mu^{2}+\lambda^{2}+12\xib\big).
\end{equation}
The final results of our evaluation of the unregularized in-vacuum expectation values of the diagonal spacelike components are
\begin{flalign}\label{vinemt11}
&\big\langle\vin\big|T_{11}\big|\vin\big\rangle = \Omega^{2}(\tau)\frac{H^{4}}{8\pi^{2}}\bigg\{
\frac{\Lambda^{4}}{3}+\Big(2\xib-\frac{\mu^{2}}{3}+\frac{14\lambda^{2}}{15}\Big)\Lambda^{2}
+\Big(\frac{\mu^{4}}{2}+6\xib\mu^{2}-\frac{\lambda^{2}}{6}\Big)\log\big(2\Lambda\big)-54\xib^{2}
-\frac{26\lambda^{4}}{105}-\frac{7\mu^{4}}{24}-8\xib\mu^{2} \nn\\
&-\frac{18\xib}{5}\lambda^{2}-\frac{\mu^{2}}{4}+\frac{19\lambda^{2}}{72}+\frac{\lambda^{2}\mu^{2}}{30}
-\frac{\gamma}{24\pi\lambda^{2}}\bigg[\frac{15}{\pi^{2}}\Big(105\pi^{-2}-15-132\xib+35\lambda^{2}-11\mu^{2}\Big)
-66\lambda^{2}-336\xib\lambda^{2}-4\lambda^{4} \nn\\
&-46\lambda^{2}\mu^{2}\bigg]\frac{\cosh\big(2\pi\lambda\big)}{\sin\big(2\pi\gamma\big)}
+\frac{\gamma}{48\pi^{2}\lambda^{3}}\bigg[\frac{15}{\pi^{2}}\Big(105\pi^{-2}-15-132\xib+175\lambda^{2}-11\mu^{2}\Big)
-366\lambda^{2}-2976\xib\lambda^{2}+136\lambda^{4} \nn\\
&-266\lambda^{2}\mu^{2}\bigg]\frac{\sinh\big(2\pi\lambda\big)}{\sin\big(2\pi\gamma\big)}
+i\csc\big(2\pi\gamma\big)\int_{-1}^{+1}C_{1r}\bigg[\Big(e^{2\pi\lambda r}+e^{2i\pi\gamma}\Big)
\psi\Big(\frac{1}{2}+\gamma+i\lambda r\Big) \nn\\
&-\Big(e^{2\pi\lambda r}+e^{-2i\pi\gamma}\Big)\psi\Big(\frac{1}{2}-\gamma+i\lambda r\Big)\bigg]dr
-\frac{i\pi}{12}\Big(3\mu^{4}+36\xib\mu^{2}-\lambda^{2}\Big)\bigg\},&&
\end{flalign}
where the coefficient $C_{1r}$ is given by
\begin{eqnarray}\label{c1}
C_{1r}&=&\frac{35}{8}\lambda^{4}r^{6}-\frac{1}{8}\Big(70\lambda^{4}+30\lambda^{2}\mu^{2}+360\xib\lambda^{2}+25\lambda^{2}\Big)r^{4}
+\frac{1}{8}\Big(39\lambda^{4}+3\mu^{4}+30\lambda^{2}\mu^{2}+396\xib\lambda^{2}
+72\xib\mu^{2}+20\lambda^{2} \nn\\
&+&6\mu^{2}+432\xib^{2}+72\xib\Big)r^{2}-\frac{1}{8}\Big(4\lambda^{4}+4\lambda^{2}\mu^{2}+60\xib\lambda^{2}
+12\xib\mu^{2}+2\lambda^{2}+2\mu^{2}+144\xib^{2}+24\xib\Big),
\end{eqnarray}
we also have
\begin{flalign}\label{vinemt22}
&\big\langle\vin\big|T_{22}\big|\vin\big\rangle = \big\langle\vin\big|T_{33}\big|\vin\big\rangle=
\Omega^{2}(\tau)\frac{H^{4}}{8\pi^{2}}\bigg\{
\frac{\Lambda^{4}}{3}+\Big(2\xib-\frac{\mu^{2}}{3}-\frac{2\lambda^{2}}{15}\Big)\Lambda^{2}
+\Big(\frac{\mu^{4}}{2}+6\xib\mu^{2}+\frac{\lambda^{2}}{6}\Big)\log\big(2\Lambda\big)-54\xib^{2}+\frac{2\lambda^{4}}{35} \nn\\
&-\frac{7\mu^{4}}{24}-8\xib\mu^{2}+\frac{4\xib}{5}\lambda^{2}-\frac{\mu^{2}}{4}
-\frac{19\lambda^{2}}{72}+\frac{2}{5}\lambda^{2}\mu^{2}
+\frac{\gamma}{16\pi\lambda^{2}}\bigg[\frac{5}{\pi^{2}}\Big(105\pi^{-2}-15-132\xib+38\lambda^{2}-11\mu^{2}\Big)
-24\lambda^{2} \nn\\
&-144\xib\lambda^{2}\bigg]\frac{\cosh\big(2\pi\lambda\big)}{\sin\big(2\pi\gamma\big)}
-\frac{\gamma}{32\pi^{2}\lambda^{3}}\bigg[\frac{5}{\pi^{2}}\Big(105\pi^{-2}-15-132\xib+178\lambda^{2}-11\mu^{2}\Big)
-124\lambda^{2}-1024\xib\lambda^{2}+\frac{200\lambda^{4}}{3} \nn\\
&-\frac{220}{3}\lambda^{2}\mu^{2}\bigg]\frac{\sinh\big(2\pi\lambda\big)}{\sin\big(2\pi\gamma\big)}
+i\csc\big(2\pi\gamma\big)\int_{-1}^{+1}C_{2r}\bigg[\Big(e^{2\pi\lambda r}+e^{2i\pi\gamma}\Big)
\psi\Big(\frac{1}{2}+\gamma+i\lambda r\Big) \nn\\
&-\Big(e^{2\pi\lambda r}+e^{-2i\pi\gamma}\Big)\psi\Big(\frac{1}{2}-\gamma+i\lambda r\Big)\bigg]dr
-\frac{i\pi}{12}\Big(3\mu^{4}+36\xib\mu^{2}+\lambda^{2}\Big)\bigg\},&&
\end{flalign}
where the coefficient $C_{2r}$ is given by
\begin{equation}\label{c2}
C_{2r}=-\frac{C_{1r}}{2}-\frac{5}{16}\lambda^{4}r^{4}+\frac{1}{16}\Big(6\lambda^{2}-6\mu^{2}+36\xib+1\Big)\lambda^{2}r^{2}
+\frac{1}{16}\Big(3\mu^{4}+2\lambda^{2}\mu^{2}+12\xib\lambda^{2}+36\xib\mu^{2}-\lambda^{2}\Big).
\end{equation}
Plugging the resulting expression given in Eq.~(\ref{wronskia}) for the Wronskian of the mode functions $U_{\k}$ into Eq.~(\ref{vint01}) and integrating
over momentum phase space, we obtain the unregularized expression for the only nonvanishing off-diagonal components
\begin{equation}\label{vinemt01}
\big\langle\vin\big|T_{01}\big|\vin\big\rangle = \big\langle\vin\big|T_{10}\big|\vin\big\rangle
= \Omega^{2}(\tau)\frac{H^{4}\lambda}{6\pi^{2}}\Lambda^{3}.
\end{equation}
The expressions (\ref{vinemt00}), (\ref{vinemt11}), (\ref{vinemt22}), and (\ref{vinemt01}) clearly have ultraviolet divergences when $\Lambda\rightarrow\infty$. This was expected, because the expectation values in the Hadamard in-vacuum state suffer from the same ultraviolet divergence properties as Minkowski spacetime.
%%%%%%%%%%%%%%%%%%%%%%%%%%%%%%%%%%%%%%%%%%%%%%%%%%%%%%%%%%%%%%%%%%%%%%%%%%%%%%%%%%%%%%%%%%%%%%%%%%%%%%%%%%%%%%%%%%%%%%%%%%%%%%%%%%%%%%%%%%%%%%%%%%%%%%%%%%%%%%%
%%%%%%%%%%%%%%%%%%%%%%%%%%%%%%%%%%%%%%%%%%%%%%%%%%%%%%%%%%%%%%%%%%%%%%%%%%%%%%%%%%%%%%%%%%%%%%%%%%%%%%%%%%%%%%%%%%%%%%%%%%%%%%%%%%%%%%%%%%%%%%%%%%%%%%%%%%%%%%%
\subsection{\label{sec:count}Construction of the counterterms and adiabatic subtractions}
To render the in-vacuum expectation values given by Eqs.~(\ref{vinemt00}), (\ref{vinemt11}), (\ref{vinemt22}), and (\ref{vinemt01}) finite, we provide a
set of needed adiabatic counterterms. The adiabatic regularization method consists of subtracting the appropriate adiabatic counterterms from the corresponding unregularized expressions. To adjust the set of appropriate counterterms, we will treat the conformal scale factor $\Omega(\tau)$, and the electromagnetic vector potential $A_{\mu}(\tau)$ as quantities of zero adiabatic order. As pointed out by Wald \cite{Wald:1978pj}, in four dimensions, to obtain a renormalized energy-momentum tensor which is consistent with the Wald axioms, the subtraction counterterms should be expanded up to fourth adiabatic order; see also \cite{Book:Birrell,Book:Parker}. Thus, to construct the expectation value of the energy-momentum tensor with the desired physical properties, we expand the subtraction counterterms up to fourth adiabatic order. To construct the appropriate counterterms, we need an expansion for the mode functions up to fourth adiabatic order. We use the definition $z=2ik\tau$ to turn the Klein-Gordon Eq.~(\ref{whitaker}) into the convenient form
\begin{equation}\label{klein}
\frac{d^{2}\mathcal{F}_{A}}{d\tau^{2}}+\Big(\omega_{0}^{2}(\tau)+\Delta(\tau)\Big)\mathcal{F}_{A}=0,
\end{equation}
where $\mathcal{F}_{A}$ is the positive frequency adiabatic solution and the conformal time dependent frequencies read
\begin{eqnarray}
\omega_{0}(\tau) &=& \Big(k^{2}+2eA_{1}kr+e^{2}A_{1}^{2}+m^{2}\Omega^{2}\Big)^{\frac{1}{2}}, \label{omega}\\
\Delta(\tau) &=& 12\xib\Big(\frac{\dot{\Omega}}{\Omega}\Big)^{2}. \label{delta}
\end{eqnarray}
It is then obvious that $\omega_{0}$ is of zero adiabatic order and $\Delta$ is of second adiabatic order. The adiabatic form of $\mathcal{F}_{A}$ is the Wentzel-Kramers-Brillouin (WKB) solution of Eq.~(\ref{klein}). This WKB solution can be written as
\begin{equation}\label{wkb}
\mathcal{F}_{A}(\tau)=\frac{1}{\sqrt{2\mathcal{W}(\tau)}}\exp\Big[-i\int^{\tau}\mathcal{W}(\tau')d\tau'\Big],
\end{equation}
where $\mathcal{W}$ satisfies the exact equation
\begin{equation}\label{exact}
\mathcal{W}^{2}=\omega_{0}^{2}+\Delta-\frac{\ddot{\mathcal{W}}}{2\mathcal{W}}+\frac{3\dot{\mathcal{W}}^{2}}{4\mathcal{W}^{2}}.
\end{equation}
To put the solution into the desired form, it is convenient to write $\mathcal{W}$ as
\begin{equation}\label{mathcalw}
\mathcal{W}=\mathcal{W}^{(0)}+\mathcal{W}^{(2)}+\mathcal{W}^{(4)},
\end{equation}
where the superscript numbers in parentheses denote the adiabatic order approximation to $\mathcal{W}$. Substituting the expansion (\ref{mathcalw}) into
Eq.~(\ref{exact}), we find the zero adiabatic order approximation
\begin{equation}\label{mathcalwz}
\mathcal{W}^{(0)}=\omega_{0}.
\end{equation}
The next iteration gives the second adiabatic order approximation
\begin{equation}\label{mathcalwt}
\mathcal{W}^{(2)}=\frac{\Delta}{2\omega_{0}}-\frac{\ddot{\omega}_{0}}{4\omega_{0}^{2}}+\frac{3\dot{\omega}_{0}^{2}}{8\omega_{0}^{3}}.
\end{equation}
Repeated iteration yields the fourth adiabatic order approximation
\begin{equation}\label{mathcalwf}
\mathcal{W}^{(4)}=-\frac{\mathcal{W}^{(2)2}}{2\omega_{0}}-\frac{\ddot{\mathcal{W}}^{(2)}}{4\omega_{0}^{2}}
+\frac{\ddot{\omega}_{0}\mathcal{W}^{(2)}}{4\omega_{0}^{3}}+\frac{3\dot{\omega}_{0}\dot{\mathcal{W}}^{(2)}}{4\omega_{0}^{3}}
-\frac{3\dot{\omega}_{0}^{2}\mathcal{W}^{(2)}}{4\omega_{0}^{4}}.
\end{equation}
It should be remarked that all terms in the adiabatic expansion of $\mathcal{W}$ of odd adiabatic order vanish. Assembling the pieces given in Eqs.~(\ref{ansatz}), (\ref{wkb}), and (\ref{mathcalw})-(\ref{mathcalwf}), we find the positive frequency adiabatic solution to fourth adiabatic order approximation
\begin{equation}\label{mathcalu}
U_{\k}^{[4]}(x)=\Omega^{-1}(\tau)\frac{1}{\sqrt{2\omega_{0}}}\bigg(1-\frac{\mathcal{W}^{(2)}}{2\omega_{0}}
-\frac{\mathcal{W}^{(4)}}{2\omega_{0}}+\frac{\mathcal{W}^{(2)2}}{2\omega_{0}^{2}}\bigg)
\exp\Big[i\k\cdot\x-i\int^{\tau}\mathcal{W}(\tau')d\tau'\Big].
\end{equation}
We use the superscript symbol $[4]$ to indicate that the cross terms from the field expansion products that are of adiabatic order greater than 4 are to
be discarded. To obtain expansions of the required counterterms to adiabatic order four, it is only necessary to calculate (\ref{vint00})-(\ref{vint01}) with $U_{\k}$ replaced by $U_{\k}^{[4]}$. Doing this, we find the counterterm to adiabatic order four for the timelike component
\begin{eqnarray}\label{delta00}
\mathcal{T}_{00}^{[4]} &=& \Omega^{2}(\tau)\frac{H^{4}}{8\pi^{2}}\bigg[
\Lambda^{4}+\Big(\mu^{2}-6\xib+\frac{2\lambda^{2}}{3}\Big)\Lambda^{2}
-\Big(\frac{\mu^{4}}{2}+6\xib\mu^{2}-\frac{\lambda^{2}}{6}\Big)\log\Big(\frac{2\Lambda}{\mu}\Big)
+72\xib^{2}-\frac{\lambda^{4}}{15}+\frac{\mu^{2}}{6}+9\xib\mu^{2}+\frac{\mu^{4}}{8} \nn\\
&-&\frac{2\lambda^{2}}{9}-2\lambda^{2}\xib-\frac{\lambda^{2}\mu^{2}}{3}-\frac{1}{60}+\frac{7\lambda^{4}}{240\mu^{4}}
+\frac{\lambda^{2}}{30\mu^{2}}-\frac{3\xib\lambda^{2}}{2\mu^{2}} \bigg].
\end{eqnarray}
We find the counterterms to adiabatic order four for the diagonal spacelike components
\begin{eqnarray}\label{delta11}
\mathcal{T}_{11}^{[4]} &=& \Omega^{2}(\tau)\frac{H^{4}}{8\pi^{2}}\bigg[\frac{\Lambda^{4}}{3}+\Big(2\xib-\frac{\mu^{2}}{3}+\frac{14\lambda^{2}}{15}\Big)\Lambda^{2}
+\Big(\frac{\mu^{4}}{2}+6\xib\mu^{2}-\frac{\lambda^{2}}{6}\Big)\log\Big(\frac{2\Lambda}{\mu}\Big)-72\xib^{2}
-\frac{13\lambda^{4}}{105}-\frac{7\mu^{4}}{24}-11\xib\mu^{2} \nn\\
&-&\frac{14\xib}{5}\lambda^{2}-\frac{\mu^{2}}{6}+\frac{7\lambda^{2}}{18}-\frac{\lambda^{2}\mu^{2}}{15}
+\frac{1}{60}-\frac{7\lambda^{4}}{240\mu^{4}}+\frac{\lambda^{2}}{18\mu^{2}}+\frac{3\xib\lambda^{2}}{2\mu^{2}} \bigg],
\end{eqnarray}
and
\begin{eqnarray}\label{delta22}
\mathcal{T}_{22}^{[4]} &=& \mathcal{T}_{33}^{[4]}=
\Omega^{2}(\tau)\frac{H^{4}}{8\pi^{2}}\bigg[
\frac{\Lambda^{4}}{3}+\Big(2\xib-\frac{\mu^{2}}{3}-\frac{2\lambda^{2}}{15}\Big)\Lambda^{2}
+\Big(\frac{\mu^{4}}{2}+6\xib\mu^{2}+\frac{\lambda^{2}}{6}\Big)\log\Big(\frac{2\Lambda}{\mu}\Big)-72\xib^{2}+\frac{\lambda^{4}}{35} \nn\\
&-&\frac{7\mu^{4}}{24}-11\xib\mu^{2}+\frac{2\xib}{5}\lambda^{2}-\frac{\mu^{2}}{6}
-\frac{7\lambda^{2}}{18}+\frac{\lambda^{2}\mu^{2}}{5}
+\frac{1}{60}+\frac{7\lambda^{4}}{720\mu^{4}}-\frac{\lambda^{2}}{45\mu^{2}}-\frac{\xib\lambda^{2}}{2\mu^{2}} \bigg].
\end{eqnarray}
We obtain the counterterms to adiabatic order four for the only nonvanishing off-diagonal components
\begin{equation}\label{delta01}
\mathcal{T}_{01}^{[4]}=\mathcal{T}_{10}^{[4]}=\Omega^{2}(\tau)\frac{H^{4}\lambda}{6\pi^{2}}\Lambda^{3}.
\end{equation}
Then, the adiabatic regularization procedure is carried out by subtracting the counterterms (\ref{delta00})-(\ref{delta01}) from the corresponding unregularized in-vacuum expectation values (\ref{vinemt00}), (\ref{vinemt11}), (\ref{vinemt22}), and (\ref{vinemt01}). Thus we obtain our final expression for the timelike component of the regularized energy-momentum tensor
\begin{flalign}\label{reg00}
T_{00}&=\big\langle\vin\big|T_{00}\big|\vin\big\rangle-\mathcal{T}_{00}^{[4]} \nn\\
&= \Omega^{2}(\tau)\frac{H^{4}}{8\pi^{2}}\bigg\{\frac{1}{60}-\frac{7\lambda^{4}}{240\mu^{4}}
-\frac{\lambda^{2}}{30\mu^{2}}+\frac{3\xib\lambda^{2}}{2\mu^{2}}-18\xib^{2}
-\frac{\lambda^{4}}{15}+\frac{\mu^{2}}{12}-3\xib\mu^{2}
-\frac{\lambda^{2}}{24}-\frac{\lambda^{2}\mu^{2}}{6} \nn\\
&+\frac{\gamma}{24\pi}\bigg(\frac{45}{\pi^{2}}-6-96\xib+4\lambda^{2}-26\mu^{2}\bigg)
\frac{\cosh\big(2\pi\lambda\big)}{\sin\big(2\pi\gamma\big)}
-\frac{\gamma}{48\pi^{2}\lambda}\bigg(\frac{45}{\pi^{2}}-6-96\xib+64\lambda^{2}-26\mu^{2}\bigg)
\frac{\sinh\big(2\pi\lambda\big)}{\sin\big(2\pi\gamma\big)} \nn\\
&+i\csc\big(2\pi\gamma\big)\int_{-1}^{+1}C_{0r}\bigg[\Big(e^{2\pi\lambda r}+e^{2i\pi\gamma}\Big)
\psi\Big(\frac{1}{2}+\gamma+i\lambda r\Big)
-\Big(e^{2\pi\lambda r}+e^{-2i\pi\gamma}\Big)\psi\Big(\frac{1}{2}-\gamma+i\lambda r\Big)\bigg]dr \nn\\
&-\Big(\frac{\mu^{4}}{2}+6\xib\mu^{2}-\frac{\lambda^{2}}{6}\Big)\log\big(\mu\big)
+\frac{i\pi}{12}\Big(3\mu^{4}+36\xib\mu^{2}-\lambda^{2}\Big)
\bigg\}.&&
\end{flalign}
We obtain our final expressions for the diagonal spacelike components of the regularized energy-momentum tensor
\begin{flalign}\label{reg11}
T_{11}&=\big\langle\vin\big|T_{11}\big|\vin\big\rangle-\mathcal{T}_{11}^{[4]} \nn\\
&=\Omega^{2}(\tau)\frac{H^{4}}{8\pi^{2}}\bigg\{
-\frac{1}{60}+\frac{7\lambda^{4}}{240\mu^{4}}-\frac{\lambda^{2}}{18\mu^{2}}-\frac{3\xib\lambda^{2}}{2\mu^{2}}
+18\xib^{2}-\frac{13\lambda^{4}}{105}+3\xib\mu^{2}-\frac{4\xib\lambda^{2}}{5}-\frac{\mu^{2}}{12}
-\frac{\lambda^{2}}{8}+\frac{\lambda^{2}\mu^{2}}{10} \nn\\
&-\frac{\gamma}{24\pi\lambda^{2}}\bigg[\frac{15}{\pi^{2}}\Big(105\pi^{-2}-15-132\xib+35\lambda^{2}-11\mu^{2}\Big)
-66\lambda^{2}-336\xib\lambda^{2}-4\lambda^{4}-46\lambda^{2}\mu^{2}\bigg]
\frac{\cosh\big(2\pi\lambda\big)}{\sin\big(2\pi\gamma\big)} \nn\\
&+\frac{\gamma}{48\pi^{2}\lambda^{3}}\bigg[\frac{15}{\pi^{2}}\Big(105\pi^{-2}-15-132\xib+175\lambda^{2}-11\mu^{2}\Big)
-366\lambda^{2}-2976\xib\lambda^{2}+136\lambda^{4}-266\lambda^{2}\mu^{2}\bigg]
\frac{\sinh\big(2\pi\lambda\big)}{\sin\big(2\pi\gamma\big)} \nn\\
&+i\csc\big(2\pi\gamma\big)\int_{-1}^{+1}C_{1r}\bigg[\Big(e^{2\pi\lambda r}+e^{2i\pi\gamma}\Big)
\psi\Big(\frac{1}{2}+\gamma+i\lambda r\Big)-\Big(e^{2\pi\lambda r}+e^{-2i\pi\gamma}\Big)
\psi\Big(\frac{1}{2}-\gamma+i\lambda r\Big)\bigg]dr \nn\\
&+\Big(\frac{\mu^{4}}{2}+6\xib\mu^{2}-\frac{\lambda^{2}}{6}\Big)\log\big(\mu\big)
-\frac{i\pi}{12}\Big(3\mu^{4}+36\xib\mu^{2}-\lambda^{2}\Big)\bigg\},&&
\end{flalign}
and
\begin{flalign}\label{reg22}
T_{22} &=T_{33} =\big\langle\vin\big|T_{33}\big|\vin\big\rangle-\mathcal{T}_{33}^{[4]} \nn\\
&=\Omega^{2}(\tau)\frac{H^{4}}{8\pi^{2}}\bigg\{
-\frac{1}{60}-\frac{7\lambda^{4}}{720\mu^{4}}+\frac{\lambda^{2}}{45\mu^{2}}+\frac{\xib\lambda^{2}}{2\mu^{2}}
+18\xib^{2}+\frac{\lambda^{4}}{35}+3\xib\mu^{2}+\frac{2\xib\lambda^{2}}{5}-\frac{\mu^{2}}{12}
+\frac{\lambda^{2}}{8}+\frac{\lambda^{2}\mu^{2}}{5} \nn\\
&+\frac{\gamma}{16\pi\lambda^{2}}\bigg[\frac{5}{\pi^{2}}\Big(105\pi^{-2}-15-132\xib+38\lambda^{2}-11\mu^{2}\Big)
-24\lambda^{2}-144\xib\lambda^{2}\bigg]\frac{\cosh\big(2\pi\lambda\big)}{\sin\big(2\pi\gamma\big)} \nn\\
&-\frac{\gamma}{32\pi^{2}\lambda^{3}}\bigg[\frac{5}{\pi^{2}}\Big(105\pi^{-2}-15-132\xib+178\lambda^{2}-11\mu^{2}\Big)
-124\lambda^{2}-1024\xib\lambda^{2}+\frac{200\lambda^{4}}{3}
-\frac{220}{3}\lambda^{2}\mu^{2}\bigg]\frac{\sinh\big(2\pi\lambda\big)}{\sin\big(2\pi\gamma\big)} \nn\\
&+i\csc\big(2\pi\gamma\big)\int_{-1}^{+1}C_{2r}\bigg[\Big(e^{2\pi\lambda r}+e^{2i\pi\gamma}\Big)
\psi\Big(\frac{1}{2}+\gamma+i\lambda r\Big)-\Big(e^{2\pi\lambda r}+e^{-2i\pi\gamma}\Big)
\psi\Big(\frac{1}{2}-\gamma+i\lambda r\Big)\bigg]dr \nn\\
&+\Big(\frac{\mu^{4}}{2}+6\xib\mu^{2}+\frac{\lambda^{2}}{6}\Big)\log\big(\mu\big)
-\frac{i\pi}{12}\Big(3\mu^{4}+36\xib\mu^{2}+\lambda^{2}\Big)\bigg\}.&&
\end{flalign}
We see that the nonvanishing unregularized expectation values of the off-diagonal components, given by Eq.~(\ref{vinemt01}), are exactly cancelled by
their counterterms (\ref{delta01}),
\begin{equation}\label{reg01}
T_{01}=T_{10}=\big\langle\vin\big|T_{10}\big|\vin\big\rangle-\mathcal{T}_{10}^{[4]}=0.
\end{equation}
Thus we have arrived at the desired expressions for the regularized in-vacuum expectation values of all the components of the energy-momentum tensor, also
called the induced energy-momentum tensor.
%%%%%%%%%%%%%%%%%%%%%%%%%%%%%%%%%%%%%%%%%%%%%%%%%%%%%%%%%%%%%%%%%%%%%%%%%%%%%%%%%%%%%%%%%%%%%%%%%%%%%%%%%%%%%%%%%%%%%%%%%%%%%%%%%%%%%%%%%%%%%%%%%%%%%%%%%%%%%%%
%%%%%%%%%%%%%%%%%%%%%%%%%%%%%%%%%%%%%%%%%%%%%%%%%%%%%%%%%%%%%%%%%%%%%%%%%%%%%%%%%%%%%%%%%%%%%%%%%%%%%%%%%%%%%%%%%%%%%%%%%%%%%%%%%%%%%%%%%%%%%%%%%%%%%%%%%%%%%%%
\section{\label{sec:prob}probing the induced energy-momentum tensor}
The nonvanishing components of the induced energy-momentum tensor are given by Eqs.~(\ref{reg00})-(\ref{reg22}). Observe that all the off-diagonal components of the induced energy-momentum tensor are zero, and the components $T_{22}$ and $T_{33}$ are equal as consequences of the underlying symmetries of the backgrounds (\ref{metric}) and (\ref{electric}). Furthermore, since the electric field background (\ref{electric}) is not invariant under full symmetries of de~Sitter spacetime, indeed violates the time reversal symmetry \cite{Antoniadis:2006wq}, and causes an electric current along its direction, we observe that $T_{00}$, $T_{11}$, and $T_{22}$ are not equal to one another. Thus we expect that in the limit of vanishing electric field background that the in-vacuum state possesses the full set of de~Sitter invariances, the induced energy-momentum tensor takes the maximally invariant form under the transformations of de~Sitter group, i.e., must be proportional to the de~Sitter metric. By setting $\lambda=0$ in Eqs.~(\ref{reg00})-(\ref{reg22}), which corresponds to vanishing electric field background, the induced energy-momentum tensor reduced to the form
\begin{equation}\label{reduced}
T_{\mu\nu}=\frac{H^{4}}{32\pi^{2}}\bigg\{\frac{1}{15}-72\xib^{2}-12\xib\mu^{2}-\frac{2\mu^{2}}{3}
+\mu^{2}\Big(12\xib+\mu^{2}\Big)\Big[\psi\Big(\frac{3}{2}+\gamma_{0}\Big)
+\psi\Big(\frac{3}{2}-\gamma_{0}\Big)-\log\big(\mu^{2}\big)\Big] \bigg\}g_{\mu\nu},
\end{equation}
where $\gamma_{0}=\sqrt{(1/4)-\mu^{2}-12\xib}$ is obtained by setting $\lambda=0$ in the definition of $\gamma$. The expression (\ref{reduced}) accords with the result of computing the renormalized vacuum energy-momentum tensor of a real scalar field in $\dsf$ obtained in Refs.~\cite{Dowker:1975tf,Bunch:1978yq}, except for the overall factor of 2 in Eq.~(\ref{reduced}). This factor 2 is consistent with the complex scalar field as being made of two real scalar fields with the number of degrees of freedom doubling up.
%%%%%%%%%%%%%%%%%%%%%%%%%%%%%%%%%%%%%%%%%%%%%%%%%%%%%%%%%%%%%%%%%%%%%%%%%%%%%%%%%%%%%%%%%%%%%%%%%%%%%%%%%%%%%%%%%%%%%%%%%%%%%%%%%%%%%%%%%%%%%%%%%%%%%%%%%%%%%%%
%%%%%%%%%%%%%%%%%%%%%%%%%%%%%%%%%%%%%%%%%%%%%%%%%%%%%%%%%%%%%%%%%%%%%%%%%%%%%%%%%%%%%%%%%%%%%%%%%%%%%%%%%%%%%%%%%%%%%%%%%%%%%%%%%%%%%%%%%%%%%%%%%%%%%%%%%%%%%%%
\subsection{\label{sec:behav}Behavior of the induced energy-momentum tensor}
\begin{figure}[t]\centering
\includegraphics[scale=0.7]{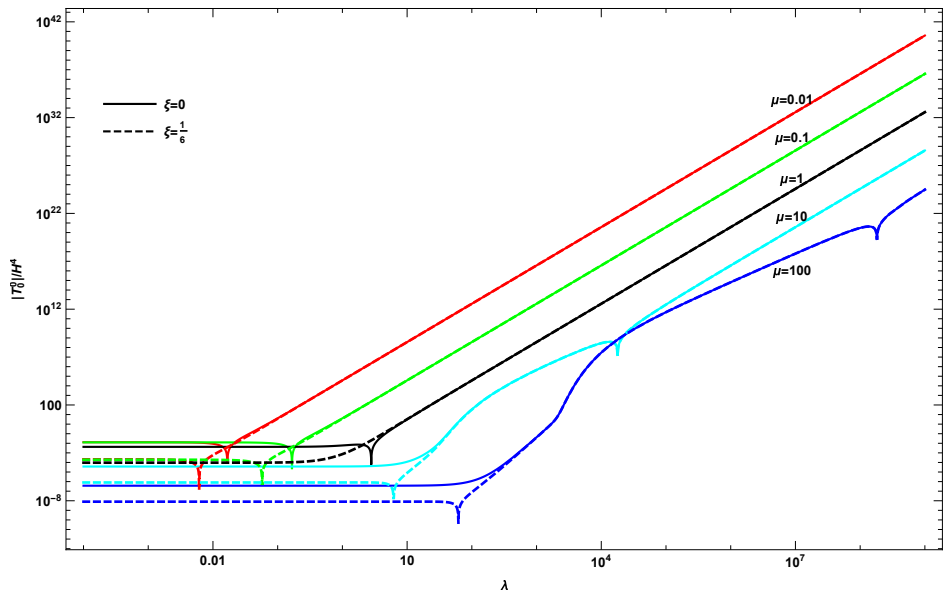}
\caption{The absolute value of $T_{0}^{0}$ component of the induced energy-momentum tensor is plotted in unit of $H^{4}$ as a function of the electric
field parameter $\lambda=-eE/H^{2}$. The graphs correspond to different values of the mass parameter $\mu=m/H$, and the coupling constant $\xi$
as indicated. Both axes have logarithmic scales.} \label{fig1}
\end{figure}
\begin{figure}[t]\centering
\includegraphics[scale=0.7]{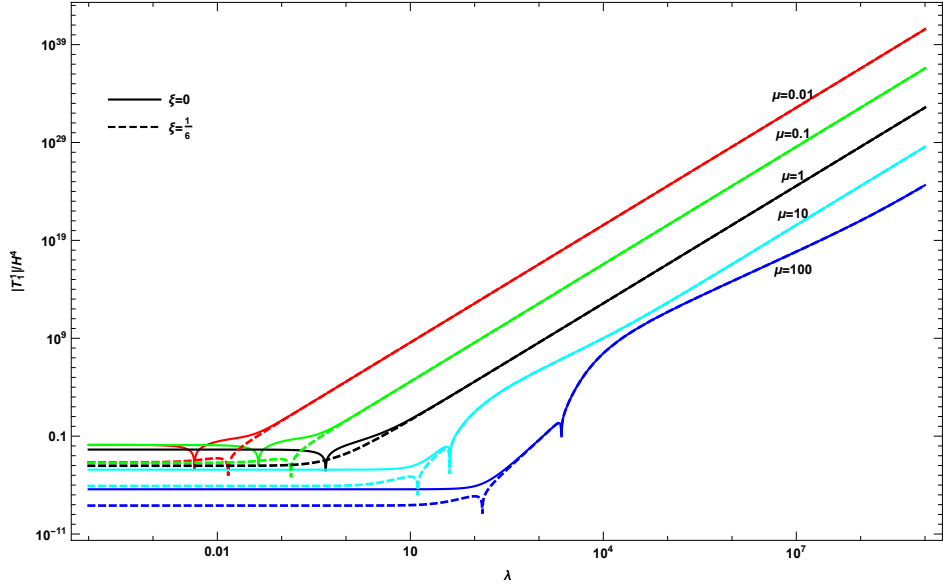}
\caption{The absolute value of $T_{1}^{1}$ component of the induced energy-momentum tensor is plotted in unit of $H^{4}$ as a function of the electric
field parameter $\lambda=-eE/H^{2}$. The graphs correspond to different values of the mass parameter $\mu=m/H$, and the coupling constant $\xi$ as indicated. Both axes have logarithmic scales.} \label{fig2}
\end{figure}

Figures~\ref{fig1}-\ref{fig3} provide some useful insight into the general behavior of the induced energy-momentum tensor. The absolute values of the expressions (\ref{reg00})-(\ref{reg22}) as functions of the electric field parameter $\lambda$, for various values of the scalar field mass parameter
$\mu$, and two values of the coupling constant $\xi$ are shown on the graphs in Figs.~\ref{fig1}-\ref{fig3}, respectively. Note especially that the scales are logarithmic on both axes, hence on the graphs zero values of the expressions, where the signs of the plots change, are displayed as singularities. Therefore, these figures signal that the induced energy-momentum tensor is analytic and varies continuously with the parameters $\lambda,\,\mu$, and $\xi$, this statement consistent with the requirements discussed in \cite{Hollands:2001nf}. Figures~\ref{fig1}-\ref{fig3} also illustrate the outstanding qualitative features of the induced energy-momentum tensor. For fixed values of $\mu$ and $\xi$, the absolute values of the nonvanishing components of induced energy-momentum tensor are increasing functions of $\lambda$, but by excluding a neighbourhood of the zero value points this behavior is assured. For fixed values of $\lambda$ and $\xi$, the absolute values of the nonvanishing components are decreasing functions of $\mu$. For fixed values of $\lambda$ and $\mu$, the nonvanishing components do not vary significantly with the parameter $\xi$ in the range $0\leq\xi\ll\lambda,\,\mu$.
The qualitative behaviors shown in Figs.~\ref{fig1}-\ref{fig3} can be given quantitative treatments by inspection of expressions (\ref{reg00})-(\ref{reg22}) in the limiting regimes. We concentrate our attention on three regimes of interest: (1) The strong electric field regime with the criterion
$\lambda\gg\max(1,\mu,\xi)$. (2) The heavy scalar field regime with the criterion $\mu\gg\max(1,\lambda,\xi)$. (3) The infrared regime with the criteria
$\mu\ll 1,\,\lambda\ll 1$, and $\xi=0$.
%%%%%%%%%%%%%%%%%%%%%%%%%%%%%%%%%%%%%%%%%%%%%%%%%%%%%%%%%%%%%%%%%%%%%%%%%%%%%%%%%%%%%%%%%%%%%%%%%%%%%%%%%%%%%%%%%%%%%%%%%%%%%%%%%%%%%%%%%%%%%%%%%%%%%%%%%%%%%%%
%%%%%%%%%%%%%%%%%%%%%%%%%%%%%%%%%%%%%%%%%%%%%%%%%%%%%%%%%%%%%%%%%%%%%%%%%%%%%%%%%%%%%%%%%%%%%%%%%%%%%%%%%%%%%%%%%%%%%%%%%%%%%%%%%%%%%%%%%%%%%%%%%%%%%%%%%%%%%%%
\subsubsection{\label{sec:stron}Strong electric field regime}
\begin{figure}[t]\centering
\includegraphics[scale=0.7]{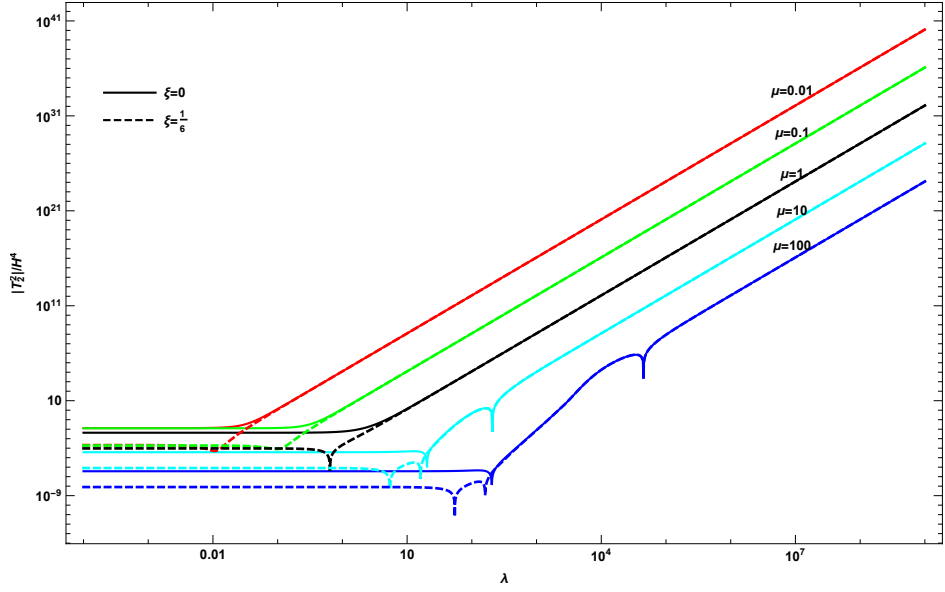}
\caption{The absolute value of $T_{2}^{2}$ component of the induced energy-momentum tensor is plotted in unit of $H^{4}$ as a function of the electric
field parameter $\lambda=-eE/H^{2}$. The graphs correspond to different values of the mass parameter $\mu=m/H$, and the coupling constant $\xi$ as indicated. Both axes have logarithmic scales. Recall that $T_{2}^{2}=T_{3}^{3}$.} \label{fig3}
\end{figure}

In the strong electric field regime $\lambda\gg\max(1,\mu,\xi)$, it is appropriate to find approximate behavior of the induced energy-momentum tensor in
the limit $\lambda\rightarrow\infty$. By expanding expressions (\ref{reg00})-(\ref{reg22}) around $\lambda=\infty$ with $\mu$ and $\xi$ fixed, we find the dominant terms in the components of the induced energy-momentum tensor
\begin{equation}\label{strong}
T_{00}=-T_{11}=3T_{22}=3T_{33}=-\Omega^{2}\frac{H^{4}}{8\pi^{2}}\Big(\frac{7\lambda^{4}}{240\mu^{4}}\Big).
\end{equation}
Thus, in this regime the absolute value of the nonvanishing components of induced energy-momentum tensor increase monotonically with increasing $\lambda$ but decrease monotonically with increasing $\mu$, as we see on the right end of any graphs in Figs.~\ref{fig1}-\ref{fig3}.
%%%%%%%%%%%%%%%%%%%%%%%%%%%%%%%%%%%%%%%%%%%%%%%%%%%%%%%%%%%%%%%%%%%%%%%%%%%%%%%%%%%%%%%%%%%%%%%%%%%%%%%%%%%%%%%%%%%%%%%%%%%%%%%%%%%%%%%%%%%%%%%%%%%%%%%%%%%%%%%
%%%%%%%%%%%%%%%%%%%%%%%%%%%%%%%%%%%%%%%%%%%%%%%%%%%%%%%%%%%%%%%%%%%%%%%%%%%%%%%%%%%%%%%%%%%%%%%%%%%%%%%%%%%%%%%%%%%%%%%%%%%%%%%%%%%%%%%%%%%%%%%%%%%%%%%%%%%%%%%
\subsubsection{\label{sec:heavy}Heavy scalar field regime}
In the heavy scalar field regime $\mu\gg\max(1,\lambda,\xi)$, it is appropriate to find approximate behavior of the induced energy-momentum tensor in the limit $\mu\rightarrow\infty$. By expanding expressions (\ref{reg00})-(\ref{reg22}) around $\mu=\infty$ with $\lambda$ and $\xi$ fixed, we find the dominant terms in the components of the induced energy-momentum tensor
\begin{eqnarray}\label{heavy}
T_{00} &=& \Omega^{2}\frac{H^{4}}{8\pi^{2}}\bigg(\frac{a}{\mu^{2}}+\frac{b}{\mu^{4}}
+\frac{c_{0}\lambda^{2}}{\mu^{4}}+\mathcal{O}(\mu^{-6})\bigg), \nn\\
T_{11} &=& -\Omega^{2}\frac{H^{4}}{8\pi^{2}}\bigg(\frac{a}{\mu^{2}}+\frac{b}{\mu^{4}}
+\frac{c_{1}\lambda^{2}}{\mu^{4}}+\mathcal{O}(\mu^{-6})\bigg), \nn\\
T_{22} &=& T_{33} = -\Omega^{2}\frac{H^{4}}{8\pi^{2}}\bigg(\frac{a}{\mu^{2}}
+\frac{b}{\mu^{4}}+\frac{c_{2}\lambda^{2}}{\mu^{4}}+\mathcal{O}(\mu^{-6})\bigg),
\end{eqnarray}
where the coefficients $a,b,c_{0},c_{1}$ and $c_{2}$ are given by
\begin{align}\label{abc}
a&=-\frac{2}{315}+\frac{\xib}{5}-72\xib^{3}, & b&=-\frac{1}{210}\Big(1-32\xib+504\xib^{2}-90720\xib^{4}\Big), \nn\\
c_{0}&= -\frac{1}{315}-\frac{7\xib}{15}+12\xib^{2}, & c_{1}&= -\frac{22}{315}+\frac{3\xib}{5}+12\xib^{2}, &
c_{2}&= \frac{2}{105}-\frac{\xib}{3}.
\end{align}
The asymptotic forms in Eq.~(\ref{heavy}) reveal that the induced energy-momentum tensor falls off as $H^{2}/m^{2}$ in the limit $(m/H)\rightarrow\infty$.
Thus, the behavior of the induced energy-momentum tensor is inconsistent with the behavior of the semiclassical energy-momentum tensor \cite{Mottola:1984ar,Bavarsad:2016cxh} that falls of as $\exp(-2\pi m/H)$ in the heavy scalar field regime where $m\gg H$. A similar feature arises for the induced electric current of both scalar \cite{Kobayashi:2014zza} and Dirac \cite{Hayashinaka:2016qqn} fields in $\dsf$. Studies \cite{Banyeres:2018aax,Hayashinaka:2018amz} have proposed explanations in physical terms for this feature of the induced electric current.
%%%%%%%%%%%%%%%%%%%%%%%%%%%%%%%%%%%%%%%%%%%%%%%%%%%%%%%%%%%%%%%%%%%%%%%%%%%%%%%%%%%%%%%%%%%%%%%%%%%%%%%%%%%%%%%%%%%%%%%%%%%%%%%%%%%%%%%%%%%%%%%%%%%%%%%%%%%%%%%
%%%%%%%%%%%%%%%%%%%%%%%%%%%%%%%%%%%%%%%%%%%%%%%%%%%%%%%%%%%%%%%%%%%%%%%%%%%%%%%%%%%%%%%%%%%%%%%%%%%%%%%%%%%%%%%%%%%%%%%%%%%%%%%%%%%%%%%%%%%%%%%%%%%%%%%%%%%%%%%
\subsubsection{\label{sec:ir}Infrared regime}
To find approximate behavior of the induced energy-momentum tensor in the infrared regime, where $\mu\ll 1,\,\lambda\ll 1$ and $\xi=0$, it is appropriate
to make Taylor series expansions of the expressions (\ref{reg00})-(\ref{reg22}) around $\mu=0$ and $\lambda=0$, and set $\xi=0$. We find that the dominant terms in the expansions of (\ref{reg00}) and (\ref{reg22}) are given by
\begin{eqnarray}
T_{00} &=&\Omega^{2}\frac{H^{4}}{8\pi^{2}}\bigg(\frac{61}{60}-\frac{17\lambda^{2}}{60\mu^{2}}
-\frac{7\lambda^{4}}{240\mu^{4}}+\mathcal{O}\Big(\lambda^{2},\mu^{2}\Big)\bigg), \label{irlw0} \\
T_{22} &=& T_{33} =-\Omega^{2}\frac{H^{4}}{8\pi^{2}}\bigg(\frac{61}{60}+\frac{11\lambda^{2}}{180\mu^{2}}
+\frac{7\lambda^{4}}{720\mu^{4}}+\mathcal{O}\Big(\lambda^{2},\mu^{2}\Big)\bigg), \label{irlw2}
\end{eqnarray}
which are valid for $\mu\ll\lambda\ll 1$ as well as $\lambda\ll\mu\ll 1$. The expansion of expression (\ref{reg11}) for $\mu\ll\lambda\ll 1$ takes the form
\begin{equation}\label{irl1}
T_{11}=\Omega^{2}\frac{H^{4}}{8\pi^{2}}\bigg(\frac{299}{60}+\frac{7\lambda^{2}}{36\mu^{2}}
+\frac{7\lambda^{4}}{240\mu^{4}}+\mathcal{O}\Big(\lambda^{2},\mu^{2}\Big)\bigg),
\end{equation}
while for $\lambda\ll\mu\ll 1$ it is approximated by
\begin{equation}\label{irw1}
T_{11}=-\Omega^{2}\frac{H^{4}}{8\pi^{2}}\bigg(\frac{61}{60}-\frac{223\lambda^{2}}{36\mu^{2}}
+\frac{1433\lambda^{4}}{240\mu^{4}}+\mathcal{O}\Big(\lambda^{2},\mu^{2}\Big)\bigg).
\end{equation}
Observe that all the asymptotic expansions (\ref{irlw0})-(\ref{irl1}) diverge as $m^{-4}$ in the exactly massless case, i.e., $m=0$. We can understand the
origin of these infrared-divergent terms by looking at the counterterms (\ref{delta00})-(\ref{delta22}). These terms arise from the contribution of the
zero modes in the massless case to the counterterms. And therefore signal that for the massless case the method of adiabatic regularization cannot be used because it leads to singularities in the counterterms, as pointed out in \cite{Parker:1974qw,Fulling:1974zr,Kobayashi:2014zza}. We emphasize that the result (\ref{irw1}) is an approximation valid only for $\lambda\ll\mu\ll 1$, and hence we cannot use it in the limit of zero mass for a fixed value of $\lambda$.
%%%%%%%%%%%%%%%%%%%%%%%%%%%%%%%%%%%%%%%%%%%%%%%%%%%%%%%%%%%%%%%%%%%%%%%%%%%%%%%%%%%%%%%%%%%%%%%%%%%%%%%%%%%%%%%%%%%%%%%%%%%%%%%%%%%%%%%%%%%%%%%%%%%%%%%%%%%%%%%
%%%%%%%%%%%%%%%%%%%%%%%%%%%%%%%%%%%%%%%%%%%%%%%%%%%%%%%%%%%%%%%%%%%%%%%%%%%%%%%%%%%%%%%%%%%%%%%%%%%%%%%%%%%%%%%%%%%%%%%%%%%%%%%%%%%%%%%%%%%%%%%%%%%%%%%%%%%%%%%
\subsection{\label{sec:trace}Trace anomaly}
It will be reassuring to do a consistency check, to see that whether the induced energy momentum tensor yields the well-known predicted trace anomaly for a free, massless, conformally invariant scalar field in $\dsf$. Putting the metric (\ref{metric}) and components (\ref{reg00})-(\ref{reg22}) together, we obtain the trace of the induced energy-momentum tensor
\begin{eqnarray}\label{trace}
T&=&g^{\mu\nu}T_{\mu\nu}=\frac{H^{4}}{8\pi^{2}}\bigg\{
\frac{1}{15}-\frac{7\lambda^{4}}{180\mu^{4}}-\frac{\lambda^{2}}{45\mu^{2}}+\frac{2\xib\lambda^{2}}{\mu^{2}}-72\xib^{2}
+\frac{\mu^{2}}{3}-12\xib\mu^{2}-\frac{\lambda^{2}}{6}-\frac{2\lambda^{2}\mu^{2}}{3}
-\frac{3\mu^{2}\gamma}{2\pi^{2}\lambda\sin\big(2\pi\gamma\big)}\nn\\
&\times&\Big(2\pi\lambda\cosh\big(2\pi\lambda\big)-\sinh\big(2\pi\lambda\big)\Big)
+\frac{i\mu^{2}}{2\sin\big(2\pi\gamma\big)}\int_{-1}^{+1}\Big(3\lambda^{2}r^{2}-\lambda^{2}-\mu^{2}-12\xib\Big)
\bigg[\Big(e^{2\pi\lambda r}+e^{2i\pi\gamma}\Big) \nn\\
&\times&\psi\Big(\frac{1}{2}+\gamma+i\lambda r\Big)
-\Big(e^{2\pi\lambda r}+e^{-2i\pi\gamma}\Big)\psi\Big(\frac{1}{2}-\gamma+i\lambda r\Big)\bigg] dr
-\mu^{2}\Big(\mu^{2}+12\xib\Big)\log\big(\mu^{2}\big)
+i\pi\mu^{2}\Big(\mu^{2}+12\xib\Big) \bigg\}.
\end{eqnarray}
We see that the trace anomaly of the free, massless, conformally coupled complex scalar field is given by
\begin{equation}\label{anomaly}
\lim_{\lambda\rightarrow 0}\lim_{\mu\rightarrow0}\lim_{\xi\rightarrow\frac{1}{6}} T= \frac{H^{4}}{120\pi^{2}}
=\frac{2}{2880\pi^{2}}\Big(\frac{1}{3}R^{2}-R_{\mu\nu}R^{\mu\nu}\Big),
\end{equation}
in arriving at the second equality, we have expressed $H^{4}$ in terms of a combination of the Ricci tensor and the scalar curvature of $\dsf$ which are given by Eq.~(\ref{ricci}), to put the result into a familiar form. The trace anomaly (\ref{anomaly}) is in agreement with the result of computing the trace anomaly \cite{Duff:1977ay} of a free, massless, conformally coupled real scalar field in $\dsf$, except for the overall factor of 2 in Eq.~(\ref{anomaly}) as explained below Eq.~(\ref{reduced}).
%%%%%%%%%%%%%%%%%%%%%%%%%%%%%%%%%%%%%%%%%%%%%%%%%%%%%%%%%%%%%%%%%%%%%%%%%%%%%%%%%%%%%%%%%%%%%%%%%%%%%%%%%%%%%%%%%%%%%%%%%%%%%%%%%%%%%%%%%%%%%%%%%%%%%%%%%%%%%%%
%%%%%%%%%%%%%%%%%%%%%%%%%%%%%%%%%%%%%%%%%%%%%%%%%%%%%%%%%%%%%%%%%%%%%%%%%%%%%%%%%%%%%%%%%%%%%%%%%%%%%%%%%%%%%%%%%%%%%%%%%%%%%%%%%%%%%%%%%%%%%%%%%%%%%%%%%%%%%%%
\section{\label{sec:curre}implications for the induced current}
The generalization of the nonconservation equation (\ref{diver:sc}) for the classical energy-momentum tensor of the scalar field to the induced energy-momentum tensor $T_{\mu\nu}$, and the induced current $j_{\mu}$ of the scalar field has important implications for the induced current. Recall that the nonvanishing components of $T_{\mu\nu}$ are given by Eqs.~(\ref{reg00})-(\ref{reg22}), and the nonzero components of $F_{\mu\nu}$ are given by Eq.~(\ref{electric}). The only nontrivial relation that arises from Eq.~(\ref{diver:sc}) is obtained by setting $\nu=0$, which leads to
\begin{equation}\label{noncons}
\partial_{0}T^{00}+H\Omega\Big(5T^{00}+T^{11}+T^{22}+T^{33}\Big)=-\Omega^{-2}Ej_{1},
\end{equation}
where we have used Eqs.~(\ref{christoff}) and (\ref{electric}). The relation (\ref{noncons}), along with the relations
\begin{align}\label{relatio}
\partial_{0}T^{00}&=-2H\Omega(\tau)T^{00}, & T&=g_{\mu\nu}T^{\mu\nu}, & -\Omega^{-2}Ej_{1}=H\Omega^{-1}A.j,
\end{align}
suffice to show that the timelike component of the induced energy-momentum tensor can be written in terms of the trace $T$, and the effective electromagnetic potential energy $A.j$ as
\begin{equation}\label{tracuur}
T_{00}=\frac{1}{4}\Omega^{2}\Big(T+A.j\Big).
\end{equation}
The induced current $j_{\mu}$ is defined as the regularized expectation value of the scalar field electric current operator (\ref{def:j}) in the in-vacuum state specified in Eq.~(\ref{vin}). It can be verified that the induced current $j_{\mu}$ is conserved and whose only non-vanishing component is $j_{1}$. Thus, the induced current flows along the electric field background direction. For convenience we write the induced current as
\begin{equation}\label{jn}
j_{\mu}=\Omega(\tau)J_{[n]} \delta_{\mu}^{1},
\end{equation}
here we use the subscript $[n]$ to indicate the adiabatic order $n$ of the subtracted counterterms to obtain the induced current $J_{[n]}$. Comparison of Eqs.~(\ref{reg00}), (\ref{trace}), and (\ref{tracuur}) allows us to read off the effective electromagnetic potential energy $A.j$, and then we will use the last relation in Eq.~(\ref{relatio}) and definition (\ref{jn}) to extract $J_{[4]}$. This gives
\begin{eqnarray}\label{j4}
J_{[4]}&=&\frac{eH^{3}}{8\pi^{2}}\bigg\{
\frac{4\xib\lambda}{\mu^{2}}-\frac{\lambda}{9\mu^{2}}-\frac{7\lambda^{3}}{90\mu^{4}}+\frac{\lambda}{3}\log\big(\mu^{2}\big)-\frac{i\pi\lambda}{3}
-\frac{4\lambda^{3}}{15}+\frac{\gamma}{6\pi\lambda}\Big(\frac{45}{\pi^{2}}-6-96\xib+4\lambda^{2}-8\mu^{2}\Big)
\frac{\cosh\big(2\pi\lambda\big)}{\sin\big(2\pi\gamma\big)} \nn\\
&-&\frac{\gamma}{12\pi^{2}\lambda^{2}}\Big(\frac{45}{\pi^{2}}-6-96\xib+64\lambda^{2}-8\mu^{2}\Big)
\frac{\sinh\big(2\pi\lambda\big)}{\sin\big(2\pi\gamma\big)}
-\frac{i\lambda}{2\sin\big(2\pi\gamma\big)}\int_{-1}^{+1}\Big(5\lambda^{2}r^{4}-\big(1+36\xib+6\lambda^{2}+3\mu^{2}\big)r^{2} \nn\\
&+&\lambda^{2}+\mu^{2}+12\xib\Big)
\bigg[\Big(e^{2\pi\lambda r}+e^{2i\pi\gamma}\Big)\psi\Big(\frac{1}{2}+\gamma+i\lambda r\Big)
-\Big(e^{2\pi\lambda r}+e^{-2i\pi\gamma}\Big)\psi\Big(\frac{1}{2}-\gamma+i\lambda r\Big)\bigg] dr \bigg\},
\end{eqnarray}
where the subscript $[4]$ indicates $J_{[4]}$ has been derived from the expressions which have been regularized by the counterterms expanded up to fourth adiabatic order. This should be clear from Eqs.~(\ref{reg00}), (\ref{tracuur}), and (\ref{jn}). To verify our result (\ref{j4}), we have evaluated directly the induced current by calculating the expectation value of the current operator (\ref{def:j}) in the in-vacuum state and then subtracting the corresponding counterterms expanded up to fourth adiabatic order. The expression for $J_{[4]}$ that follows from this direct analysis reproduces exactly the expression given by Eq.~(\ref{j4}).
\par
In Ref.~\cite{Kobayashi:2014zza}, the induced current of the massive, minimally coupled $\xi=0$, scalar field has been evaluated by calculating the expectation value of the current operator~(\ref{def:j}) in the in-vacuum state which is represented by the mode functions given in Eqs.~(\ref{umode})
and~(\ref{vmode}) with $\xi=0$ for this case. In order to regularize the expectation value, the method of adiabatic subtraction was employed. Those authors expanded the required counterterm up to second adiabatic order that it suffices to remove the divergences and the regularized expression resulting from use of it reduces to the expected results in the Minkowski spacetime limit. We refer to this prescription as minimal subtraction. Furthermore, they argued that including the contribution of fourth adiabatic order in the counterterm spoils the expected behavior in the Minkowski spacetime limit. Following these restrictions and arguments, it was subsequently found in Ref.~\cite{Kobayashi:2014zza} that the induced current is given in terms of $J_{[2]}$ as in Eq.~(\ref{jn}) by
\begin{eqnarray}\label{j2}
J_{[2]}&=&\frac{eH^{3}}{8\pi^{2}}\bigg\{
\frac{\lambda}{3}\log\big(\mu^{2}\big)-\frac{i\pi\lambda}{3}
-\frac{4\lambda^{3}}{15}+\frac{\bar{\gamma}}{6\pi\lambda}\Big(\frac{45}{\pi^{2}}+10+4\lambda^{2}-8\mu^{2}\Big)
\frac{\cosh\big(2\pi\lambda\big)}{\sin\big(2\pi\bar{\gamma}\big)}-\frac{\bar{\gamma}}{12\pi^{2}\lambda^{2}}
\Big(\frac{45}{\pi^{2}}+10+64\lambda^{2}-8\mu^{2}\Big) \nn\\
&\times&\frac{\sinh\big(2\pi\lambda\big)}{\sin\big(2\pi\bar{\gamma}\big)}
-\frac{i\lambda}{2\sin\big(2\pi\bar{\gamma}\big)}\int_{-1}^{+1}\Big(5\lambda^{2}r^{4}
+\big(5-6\lambda^{2}-3\mu^{2}\big)r^{2}+\lambda^{2}+\mu^{2}-2\Big)
\bigg[\Big(e^{2\pi\lambda r}+e^{2i\pi\bar{\gamma}}\Big) \nn\\
&\times&\psi\Big(\frac{1}{2}+\bar{\gamma}+i\lambda r\Big)
-\Big(e^{2\pi\lambda r}+e^{-2i\pi\bar{\gamma}}\Big)\psi\Big(\frac{1}{2}-\bar{\gamma}+i\lambda r\Big)\bigg] dr \bigg\},
\end{eqnarray}
where $\bar{\gamma}=\sqrt{9/4-\lambda^{2}-\mu^{2}}$ is obtained by setting $\xi=0$ in the definition of $\gamma$, and the subscript $[2]$ indicates $J_{[2]}$ has been regularized by the counterterms expanded up to second adiabatic order. It is important to note that there are two differences between expressions (\ref{j4}) and (\ref{j2}). First, the coupling constant $\xi$ is treated as arbitrary real value parameter in the expression (\ref{j4}), whereas the expression (\ref{j2}) has been computed for fixed value $\xi=0$. Second, expression (\ref{j4}) contains contributions from the fourth adiabatic order expansion of the counterterm. With these two differences, (\ref{j4}) is a generalization of (\ref{j2}). Therefore, the difference $J_{[4]}\big(\xi=0\big)-J_{[2]}$ would give only the fourth adiabatic order contribution of the counterterm to $J_{[4]}$, that is
\begin{equation}\label{j4mj2}
J_{[4]}\big(\xi=0\big)-J_{[2]}=-\frac{eH^{3}}{8\pi^{2}}\bigg(\frac{7\lambda}{9\mu^{2}}+\frac{7\lambda^{3}}{90\mu^{4}}\bigg).
\end{equation}
In our subsequent discussion, we demonstrate that the expected behavior of $J_{[4]}$ in Minkowski spacetime is not spoiled by these fourth adiabatic order contributions. Furthermore, we show that these contributions alter the behavior of the $J_{[4]}$ when compared to $J_{[2]}$, especially in the two regimes of the infrared hyperconductivity and the strong electric field.
%%%%%%%%%%%%%%%%%%%%%%%%%%%%%%%%%%%%%%%%%%%%%%%%%%%%%%%%%%%%%%%%%%%%%%%%%%%%%%%%%%%%%%%%%%%%%%%%%%%%%%%%%%%%%%%%%%%%%%%%%%%%%%%%%%%%%%%%%%%%%%%%%%%%%%%%%%%%%%%
%%%%%%%%%%%%%%%%%%%%%%%%%%%%%%%%%%%%%%%%%%%%%%%%%%%%%%%%%%%%%%%%%%%%%%%%%%%%%%%%%%%%%%%%%%%%%%%%%%%%%%%%%%%%%%%%%%%%%%%%%%%%%%%%%%%%%%%%%%%%%%%%%%%%%%%%%%%%%%%
\subsection{\label{sec:mink}Minkowski spacetime limit}
Here we explore the behavior of the result (\ref{j4}) in the Minkowski spacetime limit. Note that in the limit where the Hubble constant tends to zero, i.e., $H\rightarrow0$ we recover Minkowski spacetime. As was mentioned in the discussion of Eq.~(\ref{j4mj2}), the difference between the expressions (\ref{j4}) and (\ref{j2}) comes from the contribution of fourth adiabatic order in the adiabatic expansion of the counterterm. Comparison of the expressions (\ref{j4}) and (\ref{j2}) shows that these fourth adiabatic order contributions are
\begin{equation}\label{fourth}
\delta J=\frac{eH^{3}}{8\pi^{2}}\bigg(\frac{4\xib\lambda}{\mu^{2}}-\frac{\lambda}{9\mu^{2}}-\frac{7\lambda^{3}}{90\mu^{4}}\bigg)
=-\frac{e}{8\pi^{2}}\Big(\frac{eE}{m^{2}}\Big)\bigg(4\xib H^{3}-\frac{H^{3}}{9}-\frac{7(eE)^{2}}{90m^{2}}H\bigg).
\end{equation}
where the second equality comes from using the definitions of $\lambda$ and $\mu$ given in Eq.~(\ref{paramete}). It is clear from the second equality in Eq.~(\ref{fourth}) that all these terms vanish in the limit $H\rightarrow0$ for fixed and finite values of $E$ and $m$. This observation will automatically insure the validity of (\ref{j4}) in the Minkowski spacetime limit. If the electric field background $E$ and the scalar field mass $m$ are regarded as fixed and finite, then by taking the limit $H\rightarrow0$ of the expressions (\ref{j4}), we find
\begin{equation}\label{minkows}
\lim_{H\rightarrow0}J_{[4]}=\frac{e^{3}}{12\pi^{3}H}E^{2}e^{-\frac{\pi m^{2}}{|eE|}},
\end{equation}
which is exactly the same as the result obtained in Ref.~\cite{Kobayashi:2014zza} for the behavior of $J_{[2]}$, given by Eq.~(\ref{j2}), in the Minkowski spacetime limit. It has been argued \cite{Kobayashi:2014zza} that in the expanding dS the inverse of the Hubble constant $H^{-1}$, in fact, is equivalent to the finite time interval between switching on and off the electric field background in Minkowski spacetime. By this argument, we can see that the behavior (\ref{minkows}) of the induced current in the Minkowski spacetime limit agrees with the electric current of the charged scalar particles produced by the Schwinger mechanism in Minkowski spacetime \cite{Anderson:2013ila,Anderson:2013zia}.
\begin{figure}[t]\centering
\includegraphics[scale=0.7]{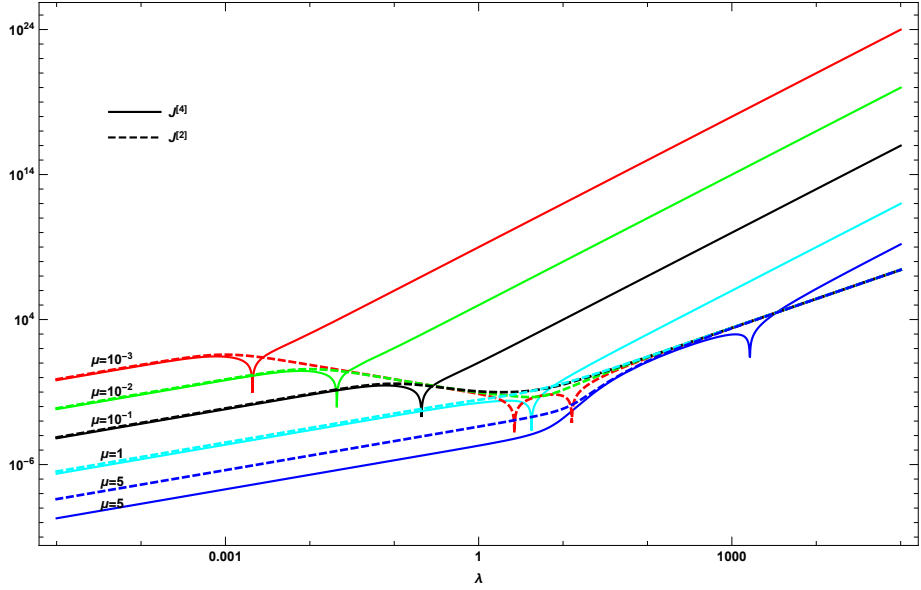}
\caption{Dependence of the absolute values of the normalized induced currents $|J_{[4]}|/eH^{3}$ (solid curves) and $|J_{[2]}|/eH^{3}$ (dashed curves) on the electric field parameter $\lambda=-eE/H^{2}$, for the case of minimal coupling $\xi=0$. The graphs correspond to different values of the scalar field mass parameter $\mu=m/H$ as indicated. Both axes have logarithmic scales.} \label{fig4}
\end{figure}
\par
A comparison of the result of this article for the induced current $J_{[4]}$ [see Eq.~(\ref{j4})] to the result of Ref.~\cite{Kobayashi:2014zza} for the induced current $J_{[2]}$ [see Eq.~(\ref{j2})] is shown in Fig.~\ref{fig4}. The figure is drawn for $\xi=0$, and illustrates that, for the light scalars $\mu<1$, the induced currents $J_{[4]}$ and $J_{[2]}$ differ considerably in the regime $\lambda\gtrsim\mu$. Subsequently, we set $\xi=0$ and concentrate our attention on the regime $\lambda\gtrsim\mu$. We examine the behaviors of $J_{[4]}$ and $J_{[2]}$ in the two regimes: The infrared hyperconductivity with the criterion $\mu<\lambda\lesssim1$, and the strong electric field $\lambda\gg\max(1,\mu,\xi)$.
%%%%%%%%%%%%%%%%%%%%%%%%%%%%%%%%%%%%%%%%%%%%%%%%%%%%%%%%%%%%%%%%%%%%%%%%%%%%%%%%%%%%%%%%%%%%%%%%%%%%%%%%%%%%%%%%%%%%%%%%%%%%%%%%%%%%%%%%%%%%%%%%%%%%%%%%%%%%%%%
%%%%%%%%%%%%%%%%%%%%%%%%%%%%%%%%%%%%%%%%%%%%%%%%%%%%%%%%%%%%%%%%%%%%%%%%%%%%%%%%%%%%%%%%%%%%%%%%%%%%%%%%%%%%%%%%%%%%%%%%%%%%%%%%%%%%%%%%%%%%%%%%%%%%%%%%%%%%%%%
\subsection{\label{sec:irhc}Behaviors in the infrared hyperconductivity regime}
In Ref.~\cite{Kobayashi:2014zza}, it was pointed out that, in the infrared hyperconductivity regime $\mu<\lambda\lesssim1$, the absolute value of the induced current $J_{[2]}$ monotonically increases with decreasing the electric field parameter $\lambda$, as shown in Fig.~\ref{fig4}. In this regime, we can approximate expression (\ref{j2}) by
\begin{equation}\label{j2:approx}
J_{[2]}\simeq\frac{eH^{3}}{8\pi^{2}}\Big(\frac{6\lambda}{\lambda^{2}+\mu^{2}}\Big).
\end{equation}
From Fig.~\ref{fig4} it is obvious that in the infrared hyperconductivity regime, the absolute value of the induced current $J_{[4]}$ reduces to zero at a certain value of the electric field parameter $\lambda_{\ast}$ which depends on the mass parameter $\mu$. Indeed, the sign of $J_{[4]}$ changes at $\lambda_{\ast}$. This figure also indicates that the absolute value of $J_{[4]}$ is a decreasing function of $\lambda$ in the interval $\mu<\lambda<\lambda_{\ast}$ and is an increasing function for $\lambda>\lambda_{\ast}$. In the infrared hyperconductivity regime, we can approximate expression (\ref{j4}) by
\begin{equation}\label{j4:approx}
J_{[4]}\simeq\frac{eH^{3}}{8\pi^{2}}\bigg(-\frac{7\lambda^{3}}{90\mu^{4}}-\frac{7\lambda}{9\mu^{2}}+\frac{6\lambda}{\lambda^{2}+\mu^{2}}\bigg).
\end{equation}
This expression has a zero at $\lambda_{\ast}\simeq2.090\mu$. In the range of $\lambda>2.090\mu$, the dominant contribution to $J_{[4]}$ comes from the first two terms in Eq.~(\ref{j4:approx}) which its absolute value increases with increasing $\lambda$. Thus, in this range, the infrared hyperconductivity phenomenon does not occur. On the other hand, in the interval $\mu<\lambda<2.090\mu$, the dominant contribution to $J_{[4]}$ comes from the last term in Eq.~(\ref{j4:approx}) which increases with decreasing $\lambda$. Thus, in the interval $\mu<\lambda<2.090\mu$, the infrared hyperconductivity phenomenon occurs. We therefore conclude that although there exist a period of the infrared hyperconductivity in our result for the induced current $J_{[4]}$, but now this phenomenon occurs in the more restricted interval $\mu<\lambda<2.090\mu$.
%%%%%%%%%%%%%%%%%%%%%%%%%%%%%%%%%%%%%%%%%%%%%%%%%%%%%%%%%%%%%%%%%%%%%%%%%%%%%%%%%%%%%%%%%%%%%%%%%%%%%%%%%%%%%%%%%%%%%%%%%%%%%%%%%%%%%%%%%%%%%%%%%%%%%%%%%%%%%%%
%%%%%%%%%%%%%%%%%%%%%%%%%%%%%%%%%%%%%%%%%%%%%%%%%%%%%%%%%%%%%%%%%%%%%%%%%%%%%%%%%%%%%%%%%%%%%%%%%%%%%%%%%%%%%%%%%%%%%%%%%%%%%%%%%%%%%%%%%%%%%%%%%%%%%%%%%%%%%%%
\subsection{\label{sec:str}Behaviors in the strong electric field regime}
Figure~\ref{fig4} shows that although the absolute values of the two induced current $J_{[2]}$ and $J_{[4]}$ are increasing functions of $\lambda$ in the strong electric field regime $\lambda\gg\max(1,\mu,\xi)$, the induced current $J_{[2]}$ does not depend on the scalar field mass, whereas the induced current $J_{[4]}$ is proportional to $\mu^{-4}$. We find that, in the limit $\lambda\rightarrow\infty$ for a fixed $\mu$, the induced current $J_{[4]}$ is given approximately by
\begin{equation}\label{j4:strong}
J_{[4]}\simeq -\frac{eH^{3}}{8\pi^{2}}\Big(\frac{7\lambda^{3}}{90\mu^{4}}\Big)=\frac{7He^{4}}{720\pi^{2}}\Big(\frac{E^{3}}{m^{4}}\Big).
\end{equation}
We observe that the behavior of the induced current $J_{[4]}$ in the Minkowski spacetime limit [see Eq.~(\ref{minkows})] will be different from that in the strong electric field limit, see Eq.~(\ref{j4:strong}). Whereas, in \cite{Kobayashi:2014zza} it was pointed out that the induced current $J_{[2]}$ has the same behavior in these two limits, indicated on the right side of Eq.~(\ref{minkows}). We note that in the Minkowski spacetime limit, the electric field strength $E$, the scalar field mass $m$, and the coupling constant $\xi$, are regarded as fixed and the Hubble constant $H$, tends to zero. Thus, in the Minkowski spacetime limit, although $\lambda=-eE/H^{2}$ and $\mu=m/H$ tend to infinity, but the ratio $\lambda/\mu^{2}$ remains finite. We also note that
in the strong electric field regime, the Hubble constant $H$, the scalar field mass $m$, and the coupling constant $\xi$, are regarded as fixed and the electric field strength $E$, tends to infinity. Thus, in this regime, $\lambda$ goes to infinity while $\mu$ is regarded as a fixed and finite value.
%%%%%%%%%%%%%%%%%%%%%%%%%%%%%%%%%%%%%%%%%%%%%%%%%%%%%%%%%%%%%%%%%%%%%%%%%%%%%%%%%%%%%%%%%%%%%%%%%%%%%%%%%%%%%%%%%%%%%%%%%%%%%%%%%%%%%%%%%%%%%%%%%%%%%%%%%%%%%%%
%%%%%%%%%%%%%%%%%%%%%%%%%%%%%%%%%%%%%%%%%%%%%%%%%%%%%%%%%%%%%%%%%%%%%%%%%%%%%%%%%%%%%%%%%%%%%%%%%%%%%%%%%%%%%%%%%%%%%%%%%%%%%%%%%%%%%%%%%%%%%%%%%%%%%%%%%%%%%%%
\section{\label{sec:concl}conclusions}
The aim of the present research was to examine the induced energy-momentum tensor of a massive, complex scalar field coupled to the electromagnetic vector potential (\ref{gauge}) which describes a uniform electric field background with a constant energy density in the conformal Poincar\'e patch of $\dsf$
where the metric takes the form (\ref{metric}). The dynamics of the complex scalar field is governed by the action (\ref{sqed}). The energy-momentum tensor of the scalar field is not covariantly conserved in the presence of the electromagnetic field, as the consequence of the electromagnetic interactions, see Eq.~(\ref{diver:sc}). The nonconservation of the scalar field energy-momentum tensor is compatible with the nonconservation of the electromagnetic field energy-momentum tensor [see Eq.~(\ref{diver:em})], and hence the total energy momentum tensor of the theory is covariantly conserved. We treated the classical gravitational field (\ref{metric}) and the classical electromagnetic field (\ref{gauge}) as fixed field configurations which they are unaffected by the dynamics of the quantum complex scalar field in response to these backgrounds. We discussed canonical quantization of the scalar field. The normalized positive and negative frequency mode functions of the scalar field which behave like the positive and negative frequency Minkowski mode functions in the remote past are given by Eqs.~(\ref{umode}) and (\ref{vmode}). These mode functions determine the in-vacuum state of the scalar field.
\par
We calculated the expectation values of all the components of the energy-momentum tensor in the in-vacuum state of the scalar field; the nonzero expectation
values have been obtained in Eqs.~(\ref{vinemt00}), (\ref{vinemt11}), (\ref{vinemt22}), and (\ref{vinemt01}). It is expected that these expectation values contain ultraviolet divergences, because the expectation values in the Hadamard in-vacuum state suffer from the same ultraviolet divergence properties as Minkowski spacetime. To render these in-vacuum expectation values finite, we employed the adiabatic regularization method. To adjust the set of appropriate counterterms, we treated the conformal scale factor and the electromagnetic vector potential as quantities of zero adiabatic order. As pointed out by Wald \cite{Wald:1978pj}, in four dimensions, to obtain a renormalized energy-momentum tensor which is consistent with the Wald axioms, the subtraction counterterms should be expanded up to fourth adiabatic order. Under these assumptions, we then constructed the set of the appropriate adiabatic counterterms, which are given by Eqs.~(\ref{delta00})-(\ref{delta01}). The adiabatic regularization procedure was carried out by subtracting the counterterms from the corresponding unregularized in-vacuum expectation values. Thus we have arrived at our final expressions for all the nonvanishing components of the induced energy-momentum tensor, which are given by Eqs.~(\ref{reg00})-(\ref{reg22}). This research has shown that all the off-diagonal components of the induced energy-momentum tensor are zero, and the components $T_{22}$ and $T_{33}$ are equal as consequences of the underlying symmetries of the backgrounds (\ref{metric}) and (\ref{electric}). Furthermore, since the electric field background (\ref{electric}) is not invariant under full symmetries of dS, indeed violates the time reversal symmetry \cite{Antoniadis:2006wq}, and causes an electric current along its direction, we observe
that $T_{00},\,T_{11}$, and $T_{22}$ are not equal to one another. The research has also shown that in the limit of zero electric field, our result for the induced energy momentum tensor reduces to the form (\ref{reduced}). The expression (\ref{reduced}) accords with the result of computing the renormalized vacuum energy-momentum tensor of a real scalar field in $\dsf$ obtained in Refs.~\cite{Dowker:1975tf,Bunch:1978yq}, except for the overall factor of 2 in Eq.~(\ref{reduced}). This factor 2 is consistent with the complex scalar field as being made of two real scalar fields with the number of degrees of freedom doubling up. The absolute values of the expressions (\ref{reg00})-(\ref{reg22}) as functions of the electric field parameter $\lambda$, for various values of the scalar field mass parameter $\mu$, and two values of the coupling constant $\xi$ are shown on the graphs in Figs.~\ref{fig1}-\ref{fig3}, respectively. These figures signal that the induced energy-momentum tensor is analytic and varies continuously with the parameters $\lambda,\,\mu$, and $\xi$, this statement consistent with the requirements discussed in \cite{Hollands:2001nf}. For fixed values of $\mu$ and $\xi$, the absolute values of the nonvanishing components of the induced energy-momentum tensor are increasing functions of $\lambda$, but by excluding a neighbourhood of the zero value points this behavior is assured. For fixed values of $\lambda$ and $\xi$, the absolute values of the nonvanishing components are decreasing functions of $\mu$. For a fixed $\lambda$ and $\mu$, the nonvanishing components do not vary significantly with the parameter $\xi$ in the range $0\leq\xi\ll\lambda,\,\mu$. The qualitative behaviors shown in Figs.~\ref{fig1}-\ref{fig3} can be given quantitative treatments by inspection of expressions (\ref{reg00})-(\ref{reg22}) in the limiting regimes. The examination of the expressions (\ref{reg00})-(\ref{reg22}) has shown that in the strong electric field regime $\lambda\gg\max(1,\mu,\xi)$, the components of the induced energy-momentum tensor can be approximated by Eq.~(\ref{strong}). In the heavy scalar field regime $\mu\gg\max(1,\lambda,\xi)$, the components can be approximated according to Eq.~(\ref{heavy}). We also found that in the infrared regime, where $\mu\ll 1,\,\lambda\ll 1,\,\xi=0$, the components can be approximated by Eqs.~(\ref{irlw0})-(\ref{irw1}). To do a consistency check, we derived the trace anomaly of the induced energy-momentum tensor for the case of a free, massless, conformally invariant scalar field in $\dsf$; see Eq.~(\ref{anomaly}). This investigation shows that our result (\ref{anomaly}) is in agreement with the earlier result of computing the trace anomaly \cite{Duff:1977ay} of a free, massless, conformally coupled real scalar field in $\dsf$, except for the overall factor of 2 in Eq.~(\ref{anomaly}), as explained below Eq.~(\ref{reduced}).
\par
One of the more significant findings to emerge from this research is that the nonconservation equation (\ref{diver:sc}) implies the relation (\ref{tracuur}) between the induced energy-momentum tensor and the induced current. This relation in turn implies the renormalization condition for the induced current. We derived the expression (\ref{j4}) for the induced current of the scalar field by using the expressions (\ref{reg00}), (\ref{trace}), and relation (\ref{tracuur}). This result for the induced current has been derived from the expressions which have been regularized by the counterterms expanded up to fourth adiabatic order. In Ref.~\cite{Kobayashi:2014zza}, the induced current of the massive, minimally coupled $\xi=0$, scalar field has been evaluated by subtracting the counterterm expanded up to second adiabatic order; see Eq.~(\ref{j2}). In the discussion of Eq.~(\ref{minkows}), we remarked that our result for the induced current in the Minkowski spacetime limit agrees with the electric current of the charged scalar particles produced by the Schwinger mechanism in Minkowski spacetime \cite{Anderson:2013ila,Anderson:2013zia}. A comparison of the result of this article for the induced current $J_{[4]}$ [see Eq.~(\ref{j4})] to the result of Ref.~\cite{Kobayashi:2014zza} for the induced current $J_{[2]}$ [see Eq.~(\ref{j2})] is shown in Fig.~\ref{fig4}. The figure is drawn for $\xi=0$, and illustrates that, for the light scalars $\mu<1$, the induced currents $J_{[4]}$ and $J_{[2]}$ differ considerably in the regime $\lambda\gtrsim\mu$. From Fig.~\ref{fig4} it is obvious that in the infrared hyperconductivity regime $\mu<\lambda\lesssim1$, the absolute value of the induced current $J_{[4]}$ reduces to zero at a certain value of the electric field parameter $\lambda_{\ast}$ which depends on the mass parameter $\mu$. We found that in the infrared hyperconductivity regime, the induced current $J_{[4]}$ can be approximated by Eq.~(\ref{j4:approx}) which has a zero at $\lambda_{\ast}\simeq2.090\mu$. In the interval $\mu<\lambda<2.090\mu$, the dominant contribution to $J_{[4]}$ comes from the last term in Eq.~(\ref{j4:approx}) which increases with decreasing $\lambda$. Thus, the infrared hyperconductivity phenomenon occurs in the interval $\mu<\lambda<2.090\mu$. We therefore conclude that although there exist a period of the infrared hyperconductivity in our result for the induced current $J_{[4]}$, but now this phenomenon occurs in the more restricted interval $\mu<\lambda<2.090\mu$. Figure~\ref{fig4} also shows that the absolute value of the induced current $J_{[4]}$ is an increasing function of $\lambda$ in the strong electric field regime $\lambda\gg\max(1,\mu,\xi)$, but decreases as $\mu^{-4}$ with increasing $\mu$. We saw that in the limit $\lambda\rightarrow\infty$ for a fixed $\mu$, the induced current $J_{[4]}$ is given approximately by Eq.~(\ref{j4:strong}). The results of this investigation show that the behavior of the induced current $J_{[4]}$ in the Minkowski spacetime limit [see Eq.~(\ref{minkows})] will be different from that in the strong electric field limit, see Eq.~(\ref{j4:strong}).
\par
This would be a fruitful area for further work. A natural progression of this work is to analyse the backreaction of the induce energy-momentum tensor on the gravitational field of $\dsf$ and the backreaction of the induce current $J_{[4]}$ on the electromagnetic field. This is also an issue for future research to explore the induced energy-momentum tensor of a Dirac field in the context of our discussion.
%%%%%%%%%%%%%%%%%%%%%%%%%%%%%%%%%%%%%%%%%%%%%%%%%%%%%%%%%%%%%%%%%%%%%%%%%%%%%%%%%%%%%%%%%%%%%%%%%%%%%%%%%%%%%%%%%%%%%%%%%%%%%%%%%%%%%%%%%%%%%%%%%%%%%%%%%%%%%%%
%%%%%%%%%%%%%%%%%%%%%%%%%%%%%%%%%%%%%%%%%%%%%%%%%%%%%%%%%%%%%%%%%%%%%%%%%%%%%%%%%%%%%%%%%%%%%%%%%%%%%%%%%%%%%%%%%%%%%%%%%%%%%%%%%%%%%%%%%%%%%%%%%%%%%%%%%%%%%%%
\begin{acknowledgments}
E.~B.~very much appreciate the support by the University of Kashan Grant No. 1143880/1.
\end{acknowledgments}
%%%%%%%%%%%%%%%%%%%%%%%%%%%%%%%%%%%%%%%%%%%%%%%%%%%%%%%%%%%%%%%%%%%%%%%%%%%%%%%%%%%%%%%%%%%%%%%%%%%%%%%%%%%%%%%%%%%%%%%%%%%%%%%%%%%%%%%%%%%%%%%%%%%%%%%%%%%%%%%
%%%%%%%%%%%%%%%%%%%%%%%%%%%%%%%%%%%%%%%%%%%%%%%%%%%%%%%%%%%%%%%%%%%%%%%%%%%%%%%%%%%%%%%%%%%%%%%%%%%%%%%%%%%%%%%%%%%%%%%%%%%%%%%%%%%%%%%%%%%%%%%%%%%%%%%%%%%%%%%
\appendix*
\section{\label{app:int}Evaluation of the momentum integrals over the mode functions}
In the appendix we present further supplementary data associated with the calculation of the expressions (\ref{vint00})-(\ref{vint22}). It is convenient
at this stage to switch to the three-dimensional spherical momentum space, and then we can perform the entire three-dimensional integrals in three-dimensional spherical coordinates. We introduce a transformation from the Cartesian momentum coordinates $(k_{x},k_{y},k_{z})$ to the spherical momentum coordinates $(k,\theta,\phi)$ by equations
\begin{align}\label{spheric}
k_{x}&=k\cos\theta, & k_{y}&=k\sin\theta\cos\phi, & k_{z}&=k\sin\theta\sin\phi,
\end{align}
In this coordinates the variable $r=k_{x}/k$ is given by $r=\cos\theta$ with range $-1\leq r\leq 1$.
The integration measure is then $d^{3}k=k^{2}\sin\theta d\phi d\theta dk$. It is useful to covert the variables of integration from the momentum $k$ to the dimensionless physical momentum $p=-k\tau$, and from the angle $\theta$ to the variable $r$. Thus, the previous expression for the integration measure may be rewritten as
\begin{align}\label{measure}
d^{3}k=-H^{3}\Omega^{3}(\tau)d\phi drp^{2}dp.
\end{align}
Accordingly, we use a dimensionless physical momentum cutoff $\Lambda$, which is related to the momentum cutoff $K$ as $\Lambda=-K\tau$. To begin the evaluation of (\ref{vint00})-(\ref{vint22}), we change the integration measure according to (\ref{measure}) and substitute the expression (\ref{umode})
for $U_{\k}(x)$. After integration over azimuth angle $\phi$ and some simplifications, the expressions (\ref{vint00})-(\ref{vint22}) reduce to
\begin{flalign}\label{reduce00}
\hspace{0.5cm}
\big\langle\vin\big|T_{00}\big|\vin\big\rangle &=
\Omega^{2}\frac{H^{4}}{8\pi^{2}}\int_{-1}^{+1}dr\bigg\{
2\mathcal{I}_{1}-4\lambda r\mathcal{I}_{2}+\Big(\frac{1}{4}-\gamma^{2}-18\xib
+\lambda^{2}r^{2}\Big)\mathcal{I}_{3}+2\Im\big[\mathcal{I}_{4}\big]-2\Re\Big[\big(6\xi-1-i\lambda r\big)\mathcal{I}_{5}\Big] \nn\\
&+\mathcal{I}_{6}\bigg\}, &&
\end{flalign}
\begin{flalign}\label{reduce11}
\hspace{0.5cm}\big\langle\vin\big|T_{11}\big|\vin\big\rangle &=
\Omega^{2}\frac{H^{4}}{8\pi^{2}}\int_{-1}^{+1}dr\bigg\{2r^{2}\mathcal{I}_{1}-4\lambda r\mathcal{I}_{2}+\Big[\big(4\xi+1\big)\lambda^{2}+\big(4\xi-1\big)\mu^{2}-\big(4\xi-1\big)\lambda^{2}r^{2} \nn\\
&+\big(6\xi-1\big)\big(8\xi-1\big)\Big]\mathcal{I}_{3}-2\Im\Big[\big(4\xi-1\big)\mathcal{I}_{4}\Big]
+2\Re\Big[\big(1-6\xi+i\lambda r-4i\lambda r\xi\big)\mathcal{I}_{5}\Big]-\big(4\xi-1\big)\mathcal{I}_{6}\bigg\},&&
\end{flalign}
and
\begin{flalign}\label{reduce22}
\hspace{0.5cm}\big\langle\vin\big|T_{22}\big|\vin\big\rangle &= \big\langle\vin\big|T_{33}\big|\vin\big\rangle
= \Omega^{2}\frac{H^{4}}{8\pi^{2}}\int_{-1}^{+1}dr\bigg\{\big(1-r^{2}\big)\mathcal{I}_{1}
+\Big[\big(4\xi-1\big)\Big(\lambda^{2}+\mu^{2}-\lambda^{2}r^{2}\Big) \nn\\
&+\big(6\xi-1\big)\big(8\xi-1\big)\Big]\mathcal{I}_{3}-2\Im\Big[\big(4\xi-1\big)\mathcal{I}_{4}\Big]
+2\Re\Big[\big(1-6\xi+i\lambda r-4i\lambda r\xi\big)\mathcal{I}_{5}\Big]-\big(4\xi-1\big)\mathcal{I}_{6}\bigg\},&&
\end{flalign}
where $\Im$ and $\Re$ stand for the imaginary and real parts of any expressions, respectively. In these expressions the momentum integrals over the Whittaker functions have been defined by
\begin{flalign}\label{i1}
\hspace{0.5cm}
\mathcal{I}_{1}=e^{\pi\lambda r}\int_{0}^{\Lambda}dpp^{3} \Big|W_{\kappa,\gamma}(-2ip)\Big|^{2},&&
\end{flalign}
\begin{flalign}\label{i2}
\hspace{0.5cm}
\mathcal{I}_{2}=e^{\pi\lambda r}\int_{0}^{\Lambda}dp p^{2} \Big|W_{\kappa,\gamma}(-2ip)\Big|^{2},&&
\end{flalign}
\begin{flalign}\label{i3}
\hspace{0.5cm}
\mathcal{I}_{3}=e^{\pi\lambda r}\int_{0}^{\Lambda}dp p \Big|W_{\kappa,\gamma}(-2ip)\Big|^{2},&&
\end{flalign}
\begin{flalign}\label{i4}
\hspace{0.5cm}
\mathcal{I}_{4}=\Big(\frac{1}{4}-\gamma^{2}-\lambda^{2}r^{2}+i\lambda r\Big)e^{\pi\lambda r}
\int_{0}^{\Lambda}dpp^{2} W_{\kappa-1,\gamma}(-2ip)W_{-\kappa,\gamma}(2ip),&&
\end{flalign}
\begin{flalign}\label{i5}
\hspace{0.5cm}
\mathcal{I}_{5}=\Big(\frac{1}{4}-\gamma^{2}-\lambda^{2}r^{2}+i\lambda r\Big)e^{\pi\lambda r}
\int_{0}^{\Lambda}dpp W_{\kappa-1,\gamma}(-2ip)W_{-\kappa,\gamma}(2ip),&&
\end{flalign}
\begin{flalign}\label{i6}
\hspace{0.5cm}
\mathcal{I}_{6}=\Big|\frac{1}{4}-\gamma^{2}-\lambda^{2}r^{2}+i\lambda r\Big|^{2}e^{\pi\lambda r}\int_{0}^{\Lambda}dpp W_{\kappa-1,\gamma}(-2ip)W_{-\kappa-1,\gamma}(2ip).&&
\end{flalign}
The integrals in Eqs.~(\ref{i1})-(\ref{i6}) are of the same kind of those momentum integrals over the Whittaker functions which encountered in the calculation of the induced current of a scalar field in two- \cite{Frob:2014zka} and four-dimensional \cite{Kobayashi:2014zza} de~Sitter spacetimes. The first step in the evaluation of the integrals (\ref{i1})-(\ref{i6}) by the method that explained in Ref.~\cite{Kobayashi:2014zza}, is to plug the Mellin-Barnes representation (\ref{mellin}) for the Whittaker function $W$ and make use of the theorem of residues. It is rather straightforward, but rather lengthy, to show that our final results are
\begin{flalign}\label{endi1}
\hspace{0.5cm}	
\mathcal{I}_{1} &=\frac{\Lambda^{4}}{4}
+\frac{\lambda r}{3}\Lambda^{3}+\frac{1}{16}\Big(4\gamma^{2}+12\lambda^{2}r^{2}-1\Big)\Lambda^{2}
+\frac{\lambda r}{8}\Big(12\gamma^{2}+20\lambda^{2}r^{2}-7\Big)\Lambda
+\frac{1}{128}\Big(27+48\gamma^{4}+560\lambda^{4}r^{4}-120\gamma^{2} \nn\\
&+480\gamma^{2}\lambda^{2}r^{2}-520\lambda^{2}r^{2}\Big)\log\big(2\Lambda\big)
-\frac{7\gamma^{4}}{32}+\frac{83\gamma^{2}}{64}-\frac{533}{96}\lambda^{4}r^{4}
+\frac{1607}{192}\lambda^{2}r^{2}-\frac{59}{16}\gamma^{2}\lambda^{2}r^{2}-\frac{351}{512}-\bigg(\frac{55\gamma^{2}}{24} \nn\\
&+\frac{35\lambda^{2} r^{2}}{8}-\frac{355}{96}\bigg)\frac{\gamma\lambda r}{\sin\big(2\pi\gamma\big)}
\Big(\cos\big(2\pi\gamma\big)+e^{2\pi\lambda r}\Big)-\frac{i}{256\sin\big(2\pi\gamma\big)}\bigg(27+48\gamma^{4}
+560\lambda^{4}r^{4}-120\gamma^{2}+480\gamma^{2}\lambda^{2}r^{2} \nn\\
&-520\lambda^{2}r^{2}\bigg)\bigg[\pi\sin\big(2\pi\gamma\big)
+\Big(e^{2\pi\lambda r}+e^{-2i\pi\gamma}\Big)\psi\Big(\frac{1}{2}-\gamma+i\lambda r\Big)
-\Big(e^{2\pi\lambda r}+e^{2i \pi\gamma}\Big)\psi\Big(\frac{1}{2}+\gamma+i\lambda r\Big)\bigg],&&
\end{flalign}
\begin{flalign}\label{endi2}
\hspace{0.5cm}	
\mathcal{I}_{2} &=\frac{\Lambda^{3}}{3}+\frac{\lambda r}{2}\Lambda^{2}
+\frac{1}{8}\Big(4\gamma^{2}+12\lambda^{2}r^{2}-1\Big)\Lambda
+\frac{\lambda r}{8}\Big(12\gamma^{2}+20\lambda^{2}r^{2}-7\Big)\log\big(2\Lambda\big)
-\frac{37}{12}\lambda^{3}r^{3}+\frac{95}{48}\lambda r-\frac{5\gamma^{2}}{4}\lambda r \nn\\
&-\Big(4\gamma{^2}+15\lambda^{2}r^{2}-4\Big)
\frac{\gamma}{6\sin\big(2\pi\gamma\big)}\Big(\cos\big(2\pi\gamma\big)+e^{2\pi\lambda r}\Big)
-\frac{i\lambda r}{16\sin\big(2\pi\gamma\big)}\Big(12\gamma^{2}+20\lambda^{2}r^{2}-7\Big) \nn\\
&\times\bigg[\pi\sin\big(2\pi\gamma\big)+\Big(e^{2\pi\lambda r}+e^{-2i\pi\gamma}\Big)
\psi\Big(\frac{1}{2}-\gamma+i\lambda r\Big)
-\Big(e^{2\pi\lambda r}+e^{2i\pi\gamma}\Big)\psi\Big(\frac{1}{2}+\gamma+i\lambda r\Big)\bigg],&&
\end{flalign}
\begin{flalign}\label{endi3}
\hspace{0.5cm}
\mathcal{I}_{3} &=\frac{\Lambda^{2}}{2}+\lambda r\Lambda+\frac{1}{8}\Big(4\gamma^{2}+12\lambda^{2}r^2-1\Big)\log\big(2\Lambda\big)
-\frac{\gamma^{2}}{4}-\frac{7\lambda^{2}r^{2}}{4}+\frac{5}{16}
-\frac{3\gamma\lambda r}{2\sin\big(2\pi\gamma\big)}\Big(\cos\big(2\pi\gamma\big)+e^{2\pi\lambda r}\Big) \nn\\
&-\frac{i}{16\sin\big(2\pi\gamma\big)}\Big(4\gamma^{2}+12\lambda^{2}r^{2}
-1\Big)\bigg[\pi\sin\big(2\pi\gamma\big)+\Big(e^{2\pi\lambda r}+e^{-2i\pi\gamma}\Big)\psi\Big(\frac{1}{2}-\gamma +i\lambda r\Big) \nn\\
&-\Big(e^{2\pi\lambda r}+e^ {2i\pi\gamma}\Big)\psi\Big(\frac{1}{2}+\gamma+i\lambda r\Big)\bigg],&&
\end{flalign}
\begin{flalign}\label{endi4}
\hspace{0.5cm}
\mathcal{I}_{4} &=-\frac{i}{16}\Big(4\gamma^{2}+4\lambda^{2}r^{2}-4i\lambda r-1\Big)\Lambda^{2}
-\frac{1}{8}\Big(4\gamma^{2}+4\lambda^{2}r^{2}-4i\lambda r-1\Big)\Big(1+2i\lambda r\Big)\Lambda
-\frac{3i}{128}\Big(4\gamma^{2}+4\lambda^{2}r^{2}-4i\lambda r \nn\\
&-1\Big)\Big(4\gamma^{2}+20\lambda^{2}r^{2}-20i\lambda r-9\Big)\log\big(2\Lambda\big)
+\frac{79}{32}i\lambda^{4}r^{4}+\frac{47}{8}\lambda^{3}r^{3}
-\frac{409}{64}i\lambda^{2}r^{2}+\frac{39\gamma^{2}}{16}i\lambda^{2}r^{2}-\frac{109}{32}\lambda r+\frac{21\gamma^{2}}{8}\lambda r \nn\\
&+\frac{7i}{32}\gamma^{4}-\frac{83i}{64}\gamma^{2}+\frac{351i}{512}
+\frac{\gamma}{4\sin\big(2\pi\gamma\big)}\Big(4\gamma^{2}+\frac{15}{2}i\lambda^{3}r^{3}+15\lambda^{2}r^{2}
+\frac{13}{2}i\gamma^{2}\lambda r-\frac{97}{8}i\lambda r-4\Big)\Big(\cos\big(2\pi\gamma\big) \nn\\
&+e^{2\pi\lambda r}\Big)-\frac{3}{256\sin\big(2\pi\gamma\big)}\Big(4\gamma^{2}+4\lambda^{2}r^{2}-4i\lambda r-1\Big)
\Big(4 \gamma ^2+20 \lambda ^2 r^2-20 i \lambda  r-9\Big)\bigg[\pi\sin\big(2\pi\gamma\big) \nn\\
&+\Big(e^{2\pi\lambda r}+e^{-2i\pi\gamma}\Big)\psi\Big(\frac{1}{2}-\gamma+i\lambda r\Big)
-\Big(e^{2\pi\lambda r}+e^{2i\pi\gamma}\Big)\psi\Big(\frac{1}{2}+\gamma+i\lambda r\Big)\bigg],&&
\end{flalign}
\begin{flalign}\label{endi5}
\hspace{0.5cm}
\mathcal{I}_{5} &=-\frac{i}{8}\Big(4\gamma^{2}+4\lambda^{2}r^{2}-4i\lambda r-1\Big)\Lambda
-\frac{1}{8}\Big(4\gamma^{2}+4\lambda^{2}r^{2}-4i\lambda r-1\Big)\Big(1+2i\lambda r\Big)\log\big(2\Lambda\big)
-2\gamma^{2}-10\lambda^{2}r^{2} \nn\\
&-\frac{16}{3}i\lambda^{3}r^{3}-4i\gamma^{2}\lambda r+\frac{20}{3}i\lambda r+\frac{3}{2}
-\frac{i\gamma}{3\sin\big(2\pi\gamma\big)}\Big(8\gamma^{2}+12\lambda^{2}r^{2}-18i\lambda r-8\Big)
\Big(\cos\big(2\pi\gamma\big)+e^{2\pi\lambda r}\Big) \nn\\
&+\frac{i}{16\sin\big(2\pi\gamma\big)}\Big(4\gamma^{2}+4\lambda^{2}r^{2}-4i\lambda r-1\Big)\Big(1+2i\lambda r\Big)
\bigg[\pi\sin\big(2\pi\gamma\big)+\Big(e^{2\pi\lambda r}+e^{-2i\pi\gamma}\Big)\psi\Big(\frac{1}{2}-\gamma+i\lambda r\Big) \nn\\
&-\Big(e^{2\pi\lambda r}+e^{2i\pi\gamma}\Big)\psi\Big(\frac{1}{2}+\gamma+i\lambda r\Big)\bigg],&&
\end{flalign}
\begin{flalign}\label{endi6}
\hspace{0.5cm}	
\mathcal{I}_{6} &=\frac{1}{64}\Big(16\gamma^{4}-8\gamma^{2}+16\lambda^{4}r^{4}
+32\gamma^{2}\lambda^{2}r^{2}+8\lambda^{2}r^{2}+1\Big)\log\big(2\Lambda\big)
-\frac{3\gamma^{4}}{16}+\frac{7\gamma^{2}}{32}-\frac{25}{48}\lambda^{4}r^{4}
-\frac{7}{8}\gamma^{2}\lambda^{2}r^{2}-\frac{29}{96}\lambda^{2}r^{2} \nn\\
&-\frac{11}{256}-\frac{\gamma}{48\sin\big(2\pi\gamma\big)}
\Big(16\gamma^{4}-8\gamma^{2}+16\lambda^{4}r^{4}+32\gamma^{2}\lambda^{2}r^{2}+8\lambda^{2}r^{2}+1\Big)
\Big(12\lambda^{3}r^{3}+20\gamma^{2}\lambda r+7\lambda r\Big)\Big(\cos\big(2\pi\gamma \big) \nn\\
&+e^{2\pi\lambda r}\Big)-\frac{i}{128\sin\big(2\pi\gamma\big)}\Big(16\gamma^{4}-8\gamma^{2}+16\lambda^{4}r^{4}
+32\gamma^{2}\lambda^{2}r^{2}+8\lambda^{2}r^{2}+1\Big)\bigg[\pi\sin\big(2\pi\gamma\big) \nn\\
&+\Big(e^{2\pi\lambda r}+e^{-2i\pi\gamma}\Big)\psi\Big(\frac{1}{2}-\gamma+i\lambda r\Big)
-\Big(e^{2\pi\lambda r}+e^{2i\pi\gamma}\Big)\psi\Big(\frac{1}{2}+\gamma+i\lambda r\Big)\bigg],&&		
\end{flalign}
where we use $\log(z)$ to denote the natural logarithm function, and $\psi(z)$ to denote the digamma function. By substituting Eqs.~(\ref{endi1})-(\ref{endi6}) into expressions (\ref{reduce00})-(\ref{reduce22}) and performing the integrals over $r$, we obtain the results
(\ref{vinemt00}), (\ref{vinemt11}), and (\ref{vinemt22}), respectively.
%%%%%%%%%%%%%%%%%%%%%%%%%%%%%%%%%%%%%%%%%%%%%%%%%%%%%%%%%%%%%%%%%%%%%%%%%%%%%%%%%%%%%%%%%%%%%%%%%%%%%%%%%%%%%%%%%%%%%%%%%%%%%%%%%%%%%%%%%%%%%%%%%%%%%%%%%%%%%%%
%%%%%%%%%%%%%%%%%%%%%%%%%%%%%%%%%%%%%%%%%%%%%%%%%%%%%%%%%%%%%%%%%%%%%%%%%%%%%%%%%%%%%%%%%%%%%%%%%%%%%%%%%%%%%%%%%%%%%%%%%%%%%%%%%%%%%%%%%%%%%%%%%%%%%%%%%%%%%%%


\begin{thebibliography}{}
\bibitem{Book:Birrell}
N.~D.~Birrell and P.~C.~W.~Davies, \textit{Quantum Fields in Curved Space} (Cambridge University Press, Cambridge, England, 1982).
%
\bibitem{Book:Parker}
L.~E.~Parker and D.~J.~Toms, \textit{Quantum Field Theory in Curved Spacetime: Quantized Fields and Gravity} (Cambridge University Press, Cambridge, England, 2009).
%
\bibitem{Book:Wald}
R.~M.~Wald, \textit{Quantum Field Theory in Curved Spacetime and Black Hole Thermodynamics} (The University of Chicago Press, Chicago, 1994).
%
\bibitem{Parker:1968mv}
L.~Parker,
%``Particle creation in expanding universes,''
Phys. Rev. Lett. \textbf{21}, 562-564 (1968). %doi:10.1103/PhysRevLett.21.562.
%
\bibitem{Parker:1969au}
L.~Parker,
%``Quantized fields and particle creation in expanding universes. I,''
Phys. Rev. \textbf{183}, 1057-1068 (1969). %doi:10.1103/PhysRev.183.1057
%
\bibitem{Parker:1971pt}
L.~Parker,
%``Quantized fields and particle creation in expanding universes. 2.,''
Phys. Rev. D \textbf{3}, 346-356 (1971) [erratum: Phys. Rev. D \textbf{3}, 2546-2546 (1971)]. %doi:10.1103/PhysRevD.3.346
%
\bibitem{Bernard:1977pq}
C.~Bernard and A.~Duncan,
%``Regularization and Renormalization of Quantum Field Theory in Curved Space-Time,''
Annals Phys. \textbf{107}, 201 (1977). %doi:10.1016/0003-4916(77)90210-X
%
\bibitem{Vilenkin:1978wc}
A.~Vilenkin,
%``Pauli-Villars Regularization and Trace Anomalies,''
Nuovo Cim. A \textbf{44}, 441-450 (1978). %doi:10.1007/BF02812985
%
\bibitem{Brown:1976wc}
Lowell~S.~Brown,
%``Stress Tensor Trace Anomaly in a Gravitational Metric: Scalar Fields,''
Phys. Rev. D \textbf{15}, 1469 (1977). %doi:10.1103/PhysRevD.15.1469
%
\bibitem{Candelas:1975du}
P.~Candelas and D.~J.~Raine,
%``General-relativistic quantum field theory: An exactly soluble model,''
Phys. Rev. D \textbf{12}, 965-974 (1975). %doi:10.1103/PhysRevD.12.965
%
\bibitem{Dowker:1975tf}
J.~S.~Dowker and R.~Critchley,
%``Effective Lagrangian and energy-momentum tensor in de Sitter space,''
Phys. Rev. D \textbf{13}, 3224 (1976). %doi:10.1103/PhysRevD.13.3224
%
\bibitem{Dowker:1976zf}
J.~S.~Dowker and R.~Critchley,
%``Stress-tensor conformal anomaly for scalar, spinor, and vector fields,''
Phys. Rev. D \textbf{16}, 3390 (1977). %doi:10.1103/PhysRevD.16.3390
%
\bibitem{Hawking:1976ja}
S.~W.~Hawking,
%``Zeta function regularization of path integrals in curved spacetime,''
Commun. Math. Phys. \textbf{55}, 133 (1977). %doi:10.1007/BF01626516
%
\bibitem{Moretti:1997qn}
V.~Moretti,
%``Direct zeta-function approach and renormalization of one-loop stress tensors in curved spacetimes,''
Phys. Rev. D \textbf{56}, 7797-7819 (1997) [arXiv:hep-th/9705060 [hep-th]]. %doi:10.1103/PhysRevD.56.7797
%
\bibitem{Christensen:1976vb}
S.~M.~Christensen,
%``Vacuum expectation value of the stress tensor in an arbitrary curved background: The covariant point-separation method,''
Phys. Rev. D \textbf{14}, 2490-2501 (1976). %doi:10.1103/PhysRevD.14.2490
%
\bibitem{Adler:1976jx}
S.~L.~Adler, J.~Lieberman and Y.~J.~Ng,
%``Regularization of the stress-energy tensor for vector and scalar particles propagating in a general background metric,''
Annals Phys. \textbf{106}, 279 (1977). %doi:10.1016/0003-4916(77)90313-X
%
\bibitem{Christensen:1978yd}
S.~M.~Christensen,
%``Regularization, renormalization, and covariant geodesic point separation,''
Phys. Rev. D \textbf{17}, 946-963 (1978). %doi:10.1103/PhysRevD.17.946
%
\bibitem{Wald:1978pj}
R.~M.~Wald,
%``Trace anomaly of a conformally invariant quantum field in curved spacetime,''
Phys. Rev. D \textbf{17}, 1477-1484 (1978). %doi:10.1103/PhysRevD.17.1477
%
\bibitem{Parker:1974qw}
L.~Parker and S.~A.~Fulling,
%``Adiabatic regularization of the energy momentum tensor of a quantized field in homogeneous spaces,''
Phys. Rev. D \textbf{9}, 341-354 (1974). %doi:10.1103/PhysRevD.9.341
%
\bibitem{Fulling:1974zr}
S.~A.~Fulling and L.~Parker,
%``Renormalization in the theory of a quantized scalar field interacting with a robertson-walker spacetime,''
Annals Phys. \textbf{87}, 176-204 (1974). %doi:10.1016/0003-4916(74)90451-5
%
\bibitem{Fulling:1974pu}
S.~A.~Fulling, L.~Parker, and B.~L.~Hu,
%``Conformal energy-momentum tensor in curved spacetime: Adiabatic regularization and renormalization,''
Phys. Rev. D \textbf{10}, 3905-3924 (1974); \textbf{11}, 1714(E) (1975). %doi:10.1103/PhysRevD.10.3905
%
\bibitem{Birrell:1978ap}
N.~D.~Birrell,
%``The application of adiabatic regularization to calculations of cosmological interest,''
Proc. Roy. Soc. Lond. A \textbf{361}, 513-526 (1978). %doi.org/10.1098/rspa.1978.0114
%
\bibitem{Hu:1978ap}
B.~L.~Hu,
%``Calculation of the trace anomaly of the conformal energy-momentum tensor in Kasner spacetime by adiabatic regularization,''
Phys. Rev. D \textbf{18}, 4460-4470 (1978). %doi:10.1103/PhysRevD.18.4460
%
\bibitem{Wald:1977up}
R.~M.~Wald,
%``The Back Reaction Effect in Particle Creation in Curved Spacetime,''
Commun. Math. Phys. \textbf{54}, 1-19 (1977). %doi:10.1007/BF01609833
%
\bibitem{Bunch:1978aq}
T.~S.~Bunch, S.~M.~Christensen, and S.~A.~Fulling,
%``Massive quantum field theory in two-dimensional Robertson-Walker space-time,''
Phys. Rev. D \textbf{18}, 4435-4459 (1978). %doi:10.1103/PhysRevD.18.4435
%
\bibitem{Bunch:1977sq}
T.~S.~Bunch and P.~C.~W.~Davies,
%``Covariant point-splitting regularization for a scalar quantum field in a Robertson-Walker universe with spatial curvature,''
Proc. Roy. Soc. Lond. A \textbf{357}, 381-394 (1977). %doi:10.1098/rspa.1977.0174
%
\bibitem{Bunch:1978yw}
T.~S.~Bunch and P.~C.~W.~Davies,
%``Non-conformal renormalised stress tensors in Robertson-Walker space-times,''
J. Phys. A \textbf{11}, 1315-1328 (1978). %doi:10.1088/0305-4470/11/7/018
%
\bibitem{Davies:1977ze}
P.~C.~W.~Davies, S.~A.~Fulling, S.~M.~Christensen and T.~S.~Bunch,
%``Energy-momentum tensor of a massless scalar quantum field in a Robertson-Walker universe,''
Annals Phys. \textbf{109}, 108-142 (1977). %doi:10.1016/0003-4916(77)90167-1
%
\bibitem{Bunch:1978gb}
T.~S.~Bunch,
%``Calculation of the renormalised quantum stress tensor by adiabatic regularisation in two- and four-dimensional Robertson-Walker space-times,''
J. Phys. A \textbf{11}, 603-607 (1978). %doi:10.1088/0305-4470/11/3/021
%
\bibitem{Bunch:1980vc}
T.~S.~Bunch,
%``Adiabatic regularisation for scalar fields with arbitrary coupling to the scalar curvature,''
J. Phys. A \textbf{13}, 1297-1310 (1980). %doi:10.1088/0305-4470/13/4/022
%
\bibitem{Anderson:1987yt}
P.~R.~Anderson and L.~Parker,
%``Adiabatic regularization in closed Robertson-Walker universes,''
Phys. Rev. D \textbf{36}, 2963 (1987). %doi:10.1103/PhysRevD.36.2963
%
\bibitem{Habib:1999cs}
S.~Habib, C.~Molina-Par\'{i}s and E.~Mottola,
%``Energy-momentum tensor of particles created in an expanding universe,''
Phys. Rev. D \textbf{61}, 024010 (1999) %doi:10.1103/PhysRevD.61.024010
[arXiv:gr-qc/9906120 [gr-qc]].
%
\bibitem{Zhang:2019gtg}
Y.~Zhang, B.~Wang and X.~Ye,
%``A massless scalar field in Robertson-Walker spacetimes: Adiabatic regularization and Green\textquoteright{}s function,''
Chin. Phys. C \textbf{44}, no.9, 095104 (2020) %doi:10.1088/1674-1137/44/9/095104
[arXiv:1909.13010 [gr-qc]].
%
\bibitem{Markkanen:2017rvi}
T.~Markkanen,
%``Renormalization of the inflationary perturbations revisited,''
J. Cosmol. Astropart. Phys. \textbf{05} (2018) 001 %doi:10.1088/1475-7516/2018/05/001
[arXiv:1712.02372 [hep-th]].
%
\bibitem{SolaPeracaula:2022hpd}
J.~Sol\`{a} Peracaula,
%``The cosmological constant problem and running vacuum in the expanding universe,''
Phil. Trans. Roy. Soc. Lond. A \textbf{380}, 20210182 (2022) %doi:10.1098/rsta.2021.0182
[arXiv:2203.13757 [gr-qc]].
%
\bibitem{Moreno-Pulido:2022phq}
C.~Moreno-Pulido and J.~Sol\`{a} Peracaula,
%``Renormalizing the vacuum energy in cosmological spacetime: implications for the cosmological constant problem,''
Eur. Phys. J. C \textbf{82}, no.6, 551 (2022) %doi:10.1140/epjc/s10052-022-10484-w
[arXiv:2201.05827 [gr-qc]].
%
\bibitem{Gibbons:1977mu}
G.~W.~Gibbons and S.~W.~Hawking,
%``Cosmological Event Horizons, Thermodynamics, and Particle Creation,''
Phys. Rev. D \textbf{15}, 2738-2751 (1977). %doi:10.1103/PhysRevD.15.2738
%
\bibitem{Bunch:1978yq}
T.~S.~Bunch and P.~C.~W.~Davies,
%``Quantum field theory in de Sitter space: renormalization by point-splitting,''
Proc. Roy. Soc. Lond. A \textbf{360}, 117-134 (1978). %doi:10.1098/rspa.1978.0060
%
\bibitem{Mottola:1984ar}
E.~Mottola,
%``Particle creation in de Sitter space,''
Phys. Rev. D \textbf{31}, 754 (1985). %doi:10.1103/PhysRevD.31.754
%
\bibitem{Anderson:2013ila}
P.~R.~Anderson and E.~Mottola,
%``Instability of global de Sitter space to particle creation,''
Phys. Rev. D \textbf{89}, 104038 (2014) %doi:10.1103/PhysRevD.89.104038
[arXiv:1310.0030 [gr-qc]].
%
\bibitem{Anderson:2013zia}
P.~R.~Anderson and E.~Mottola,
%``Quantum vacuum instability of \textquotedblleft{}eternal\textquotedblright{} de Sitter space,''
Phys. Rev. D \textbf{89}, 104039 (2014) %doi:10.1103/PhysRevD.89.104039
[arXiv:1310.1963 [gr-qc]].
%
\bibitem{Markkanen:2016aes}
T.~Markkanen and A.~Rajantie,
%``Massive scalar field evolution in de Sitter,''
J. High Energy Phys. \textbf{01} (2017) 133  %doi:10.1007/JHEP01(2017)133
[arXiv:1607.00334 [gr-qc]].
%
\bibitem{Anderson:2000wx}
P.~R.~Anderson, W.~Eaker, S.~Habib, C.~Molina-Par\'{i}s and E.~Mottola,
%``Attractor states and infrared scaling in de Sitter space,''
Phys. Rev. D \textbf{62}, 124019 (2000) %doi:10.1103/PhysRevD.62.124019
[arXiv:gr-qc/0005102 [gr-qc]].
%
\bibitem{Zhang:2019urk}
Y.~Zhang, X.~Ye and B.~Wang,
%``Adiabatic regularization and Green\textquoteright{}s function of a scalar field in de Sitter space: Positive energy spectrum and no trace anomaly,''
Sci. China Phys. Mech. Astron. \textbf{63}, no.5, 250411 (2020) %doi:10.1007/s11433-019-1451-1
[arXiv: 1903.10115 [gr-qc]].
%
\bibitem{Ye:2022tgs}
X.~Ye, Y.~Zhang and B.~Wang,
%``Point-splitting regularization of the stress tensor of a coupling scalar field in de Sitter space,''
J. Cosmol. Astropart. Phys. \textbf{09} (2022) 020 %doi:10.1088/1475-7516/2022/09/020
[arXiv:2205.04761 [gr-qc]].
%
\bibitem{Schwinger:1951nm}
J.~Schwinger,
%``On gauge invariance and vacuum polarization,''
Phys. Rev. \textbf{82}, 664-679 (1951). %doi:10.1103/PhysRev.82.664
%
\bibitem{Heisenberg:1936eu}
W.~Heisenberg and H.~Euler
%``Consequences of Dirac\textquoteright{}s Theory of the Positron,''
Z. Phys. \textbf{98}, 714-732 (1936) [arXiv:physics/0605038 [physics.hist-ph]].
%
\bibitem{Sauter:1931zz}
F.~Sauter,
%``\"{U}ber das Verhalten eines Elektrons im homogenen elektrischen Feld nach der relativistischen Theorie Diracs,''
Z. Phys. \textbf{69}, 742-764 (1931). %doi:10.1007/BF01339461
%
\bibitem{Gelis:2015kya}
F.~Gelis and N.~Tanji,
%``Schwinger mechanism revisited,''
Prog. Part. Nucl. Phys. \textbf{87}, 1-49 (2016) %doi:10.1016/j.ppnp.2015.11.001
[arXiv:1510.05451 [hep-ph]].
%
\bibitem{Dunne:2004nc}
G.~V.~Dunne,
``\textit{Heisenberg-Euler effective Lagrangians: Basics and extensions},''
edited by M.~Shifman \textit{et al.}, From Fields to Strings (World Scientific Publishing, Singapore, 2005), Vol. 1, pp. 445–522. %doi:10.1142/9789812775344\_0014
[arXiv:hep-th/0406216 [hep-th]].
%
\bibitem{Antoniadis:2006wq}
I.~Antoniadis, P.~O.~Mazur and E.~Mottola,
%``Cosmological dark energy: prospects for a dynamical theory,''
New J. Phys. \textbf{9}, 11 (2007) %doi:10.1088/1367-2630/9/1/011
[arXiv:gr-qc/0612068 [gr-qc]].
%
\bibitem{Martin:2007bw}
J.~Martin,
%``Inflationary Perturbations: The Cosmological Schwinger Effect,''
Lect. Notes Phys. \textbf{738}, 193-241 (2008) %doi:10.1007/978-3-540-74353-8\_6
[arXiv:0704.3540 [hep-th]].
%
\bibitem{Maleknejad:2012fw}
A.~Maleknejad, M.~M.~Sheikh-Jabbari and J.~Soda,
%``Gauge fields and inflation,''
Phys. Rept. \textbf{528}, 161-261 (2013) %doi:10.1016/j.physrep.2013.03.003
[arXiv:1212.2921 [hep-th]].
%
\bibitem{Durrer:2013pga}
R.~Durrer and A.~Neronov,
%``Cosmological magnetic fields: their generation, evolution and observation,''
Astron. Astrophys. Rev. \textbf{21}, 62 (2013) %doi:10.1007/s00159-013-0062-7
[arXiv:1303.7121 [astro-ph.CO]].
%
\bibitem{Garriga:1993fh}
J.~Garriga,
%``Nucleation rates in flat and curved space,''
Phys. Rev. D \textbf{49}, 6327-6342 (1994) %doi:10.1103/PhysRevD.49.6327
[arXiv:hep-ph/9308280 [hep-ph]].
%
\bibitem{Garriga:1994bm}
J.~Garriga,
%``Pair production by an electric field in (1+1)-dimensional de Sitter space,''
Phys. Rev. D \textbf{49}, 6343-6346 (1994). %doi:10.1103/PhysRevD.49.6343
%
\bibitem{Kim:2008xv}
S.~P.~Kim and D.~N.~Page,
%``Schwinger Pair Production in $dS_{2}$ and $AdS_{2}$,''
Phys. Rev. D \textbf{78}, 103517 (2008) %doi:10.1103/PhysRevD.78.103517
[arXiv:0803.2555 [hep-th]].
%
\bibitem{Cai:2014qba}
R.~G.~Cai and S.~P.~Kim,
%``One-loop effective action and Schwinger effect in (anti-) de Sitter space,''
J. High Energy Phys. \textbf{09} (2014) 072 %doi:10.1007/JHEP09(2014)072
[arXiv:1407.4569 [hep-th]].
%
\bibitem{Hamil:2018rvu}
B.~Hamil and M.~Merad,
%``Schwinger mechanism on de Sitter background,''
Int. J. Mod. Phys. A \textbf{33}, no.30, 1850177 (2018). %doi:10.1142/S0217751X18501774
%
\bibitem{Frob:2014zka}
M.~B.~Fr\"{o}b, J.~Garriga, S.~Kanno, M.~Sasaki, J.~Soda, T.~Tanaka and A.~Vilenkin,
%``Schwinger effect in de Sitter space,''
J. Cosmol. Astropart. Phys. \textbf{04} (2014) 009 %doi:10.1088/1475-7516/2014/04/009
[arXiv:1401.4137 [hep-th]].
%
\bibitem{Kobayashi:2014zza}
T.~Kobayashi and N.~Afshordi,
%``Schwinger effect in 4D de Sitter space and constraints on magnetogenesis in the early universe,''
J. High Energy Phys. \textbf{10} (2014) 166 %doi:10.1007/JHEP10(2014)166
[arXiv:1408.4141 [hep-th]].
%
\bibitem{Bavarsad:2016cxh}
E.~Bavarsad, C.~Stahl and S.~S.~Xue,
%``Scalar current of created pairs by Schwinger mechanism in de Sitter spacetime,''
Phys. Rev. D \textbf{94}, no.10, 104011 (2016) %doi:10.1103/PhysRevD.94.104011
[arXiv:1602.06556 [hep-th]].
%
\bibitem{Villalba:1995za}
V.~M.~Villalba,
%``Creation of spin-1/2 particles by an electric field in de Sitter space,''
Phys. Rev. D \textbf{52}, 3742-3745 (1995) %doi:10.1103/PhysRevD.52.3742
[arXiv:hep-th/9507021 [hep-th]].
%
\bibitem{Haouat:2012ik}
S.~Haouat and R.~Chekireb,
%``Comment on \textquotedblleft{}Creation of spin 1/2 particles by an electric field in de~Sitter space\textquotedblright{},''
Phys. Rev. D \textbf{87}, no.8, 088501 (2013) %doi:10.1103/PhysRevD.87.088501
[arXiv:1207.4342 [hep-th]].
%
\bibitem{Stahl:2015gaa}
C.~Stahl, E.~Strobel and S.~S.~Xue,
%``Fermionic current and Schwinger effect in de Sitter spacetime,''
Phys. Rev. D \textbf{93}, no.2, 025004 (2016) %doi:10.1103/PhysRevD.93.025004
[arXiv:1507.01686 [gr-qc]].
%
\bibitem{Haouat:2015uaa}
S.~Haouat and R.~Chekireb,
%``Effect of the electric field on the creation of fermions in de-Sitter space-time,''
[arXiv:1504.08201 [gr-qc]].
%
\bibitem{Stahl:2015cra}
C.~Stahl and S.~Eckhard,
%``Semiclassical fermion pair creation in de Sitter spacetime,''
AIP Conf. Proc. \textbf{1693}, no.1, 050005 (2015) %doi:10.1063/1.4937198
[arXiv:1507.01401 [hep-th]].
%
\bibitem{Hayashinaka:2016dnt}
T.~Hayashinaka and J.~Yokoyama,
%``Point splitting renormalization of Schwinger induced current in de~Sitter spacetime,''
J. Cosmol. Astropart. Phys. \textbf{07} (2016) 012 %doi:10.1088/1475-7516/2016/07/012
[arXiv:1603.06172 [hep-th]].
%
\bibitem{Geng:2017zad}
J.~J.~Geng, B.~F.~Li, J.~Soda, A.~Wang, Q.~Wu and T.~Zhu,
%``Schwinger pair production by electric field coupled to inflaton,''
J. Cosmol. Astropart. Phys. \textbf{02} (2018) 018 %doi:10.1088/1475-7516/2018/02/018
[arXiv:1706.02833 [gr-qc]].
%
\bibitem{Bavarsad:2017oyv}
E.~Bavarsad, S.~P.~Kim, C.~Stahl and S.~S.~Xue,
%``Effect of a magnetic field on Schwinger mechanism in de Sitter spacetime,''
Phys. Rev. D \textbf{97}, no.2, 025017 (2018) %doi:10.1103/PhysRevD.97.025017
[arXiv:1707.03975 [hep-th]].
%
\bibitem{Bavarsad:2018lvn}
E.~Bavarsad, S.~P.~Kim, C.~Stahl and S.~S.~Xue,
%``Schwinger mechanism in electromagnetic field in de~Sitter spacetime,''
Eur. Phys. J. Web Conf. \textbf{168}, 03002 (2018). %doi:10.1051/epjconf/201816803002
%
\bibitem{Moradi:2009zz}
S.~Moradi,
%``Particle production in cosmological spacetimes with electromagnetic fields,''
Mod. Phys. Lett. A \textbf{24}, 1129-1136 (2009). %doi:10.1142/S0217732309028801
%
\bibitem{Hayashinaka:2016qqn}
T.~Hayashinaka, T.~Fujita and J.~Yokoyama,
%``Fermionic Schwinger effect and induced current in de~Sitter space,''
J. Cosmol. Astropart. Phys. \textbf{07} (2016) 010 %doi:10.1088/1475-7516/2016/07/010
[arXiv:1603.04165 [hep-th]].
%
\bibitem{Banyeres:2018aax}
M.~Banyeres, G.~Dom\`{e}nech and J.~Garriga,
%``Vacuum birefringence and the Schwinger effect in (3+1) de~Sitter,''
J. Cosmol. Astropart. Phys. \textbf{10} (2018) 023 %doi:10.1088/1475-7516/2018/10/023
[arXiv:1809.08977 [hep-th]].
%
\bibitem{Hayashinaka:2018amz}
T.~Hayashinaka and S.~S.~Xue,
%``Physical renormalization condition for de Sitter QED,''
Phys. Rev. D \textbf{97}, no.10, 105010 (2018) %doi:10.1103/PhysRevD.97.105010
[arXiv:1802.03686 [gr-qc]].
%
\bibitem{AkbariAhmadmahmoudi:2021tpj}
M.~Akbari Ahmadmahmoudi and E.~Bavarsad,
%``Energy-momentum tensor and effective Lagrangian of scalar QED with a nonminimal coupling in 2D de Sitter spacetime,''
Phys. Rev. D \textbf{103}, no.10, 105009 (2021) %doi:10.1103/PhysRevD.103.105009
[arXiv:2102.03407 [hep-th]].
%
\bibitem{Bavarsad:2018mor}
E.~Bavarsad and M.~Mortezazadeh,
%``Trace of energy-momentum tensor and gravitational backreaction of Schwinger scalars in 3D de Sitter spacetime,''
Iran. J. Phys. Res. \textbf{18}, 91 (2018).
%
\bibitem{Bavarsad:2019jlg}
E.~Bavarsad, S.~P.~Kim, C.~Stahl and S.~S.~Xue,
\textit{Effect of Schwinger pair production on the evolution of the Hubble constant in de~Sitter spacetime},
edited by E.~Battistelli  \textit{et al.},
the Fifteenth Marcel Grossmann Meeting (World Scientific Publishing, Singapore, 2022), pp.~2041-2046,
DOI:10.1142/9789811258251\_0306 [arXiv:1909.09319 [hep-th]].
%
\bibitem{Botshekananfard:2019zak}
M.~Botshekananfard and E.~Bavarsad,
%``Induced energy-momentum tensor of a Dirac field in 2D de~Sitter QED,''
Phys. Rev. D \textbf{101}, no.8, 085011 (2020) %doi:10.1103/PhysRevD.101.085011
[arXiv:1911.10588 [hep-th]].
%
\bibitem{Bavarsad:2017wbe}
E.~Bavarsad and N.~Margosian,
%``Gravitational backreaction effect of Schwinger pair production in a strong electric field in de~Sitter spacetime,''
J. Res. Many-Body Syst. \textbf{8}, 1 (2018), DOI: 10.22055/jrmbs.2018.13879.
%
\bibitem{Haouat:2012dr}
S.~Haouat and R.~Chekireb,
%``Effect of electromagnetic fields on the creation of scalar particles in a flat Robertson-Walker space-time,''
Eur. Phys. J. C \textbf{72}, 2034 (2012) %doi:10.1140/epjc/s10052-012-2034-x
[arXiv:1201.5738 [hep-th]].
%
\bibitem{Shakeri:2019mnt}
S.~Shakeri, M.~A.~Gorji and H.~Firouzjahi,
%``Schwinger Mechanism During Inflation,''
Phys. Rev. D \textbf{99}, no.10, 103525 (2019) %doi:10.1103/PhysRevD.99.103525
[arXiv:1903.05310 [hep-th]].
%
\bibitem{Stahl:2018idd}
C.~Stahl,
%``Schwinger effect impacting primordial magnetogenesis,''
Nucl. Phys. B \textbf{939}, 95-104 (2019) %doi:10.1016/j.nuclphysb.2018.12.017
[arXiv:1806.06692 [hep-th]].
%
\bibitem{Giovannini:2018qbq}
M.~Giovannini,
%``Spectator electric fields, de Sitter spacetime, and the Schwinger effect,''
Phys. Rev. D \textbf{97}, no.6, 061301(R) (2018) %doi:10.1103/PhysRevD.97.061301
[arXiv:1801.09995 [hep-th]].
%
\bibitem{Sharma:2017ivh}
R.~Sharma and S.~Singh,
%``Multifaceted Schwinger effect in de Sitter space,''
Phys. Rev. D \textbf{96}, no.2, 025012 (2017) %doi:10.1103/PhysRevD.96.025012
[arXiv:1704.05076 [gr-qc]].
%
\bibitem{Domcke:2019qmm}
V.~Domcke, Y.~Ema and K.~Mukaida,
%``Chiral Anomaly, Schwinger Effect, Euler-Heisenberg Lagrangian, and application to axion inflation,''
J. High Energy Phys. \textbf{02} (2020) 055 %doi:10.1007/JHEP02(2020)055
[arXiv:1910.01205 [hep-ph]].
%
\bibitem{Tangarife:2017rgl}
W.~Tangarife, K.~Tobioka, L.~Ubaldi and T.~Volansky,
%``Dynamics of Relaxed Inflation,''
J. High Energy Phys. \textbf{02} (2018) 084 %doi:10.1007/JHEP02(2018)084
[arXiv:1706.03072 [hep-ph]].
%
\bibitem{Kitamoto:2018htg}
H.~Kitamoto,
%``Schwinger Effect in Inflaton-Driven Electric Field,''
Phys. Rev. D \textbf{98}, no.10, 103512 (2018) %doi:10.1103/PhysRevD.98.103512
[arXiv:1807.03753 [hep-th]].
%
\bibitem{Chua:2018dqh}
W.~Z.~Chua, Q.~Ding, Y.~Wang and S.~Zhou,
%``Imprints of Schwinger effect on primordial spectra,''
J. High Energy Phys. \textbf{04} (2019) 066 %doi:10.1007/JHEP04(2019)066
[arXiv:1810.09815 [hep-th]].
%
\bibitem{Gorbar:2019fpj}
E.~V.~Gorbar, A.~I.~Momot, O.~O.~Sobol and S.~I.~Vilchinskii,
%``Kinetic approach to the Schwinger effect during inflation,''
Phys. Rev. D \textbf{100}, no.12, 123502 (2019) %doi:10.1103/PhysRevD.100.123502
[arXiv:1909.10332 [gr-qc]].
%
\bibitem{Sobol:2019xls}
O.~O.~Sobol, E.~V.~Gorbar and S.~I.~Vilchinskii,
%``Backreaction of electromagnetic fields and the Schwinger effect in pseudoscalar inflation magnetogenesis,''
Phys. Rev. D \textbf{100}, no.6, 063523 (2019) %doi:10.1103/PhysRevD.100.063523
[arXiv:1907.10443 [astro-ph.CO]].
%
\bibitem{Sobol:2018djj}
O.~O.~Sobol, E.~V.~Gorbar, M.~Kamarpour and S.~I.~Vilchinskii,
%``Influence of backreaction of electric fields and Schwinger effect on inflationary magnetogenesis,''
Phys. Rev. D \textbf{98}, no.6, 063534 (2018) %doi:10.1103/PhysRevD.98.063534
[arXiv:1807.09851 [hep-ph]].
%
\bibitem{Kim:2019joy}
S.~P.~Kim,
%``Astrophysics in Strong Electromagnetic Fields and Laboratory Astrophysics,''
[arXiv: 1905.13439 [gr-qc]].
%
\bibitem{Book:NIST}
F.~W.~J.~Olver, D.~W.~Lozier, R.~F.~Boisvert, and C.~W.~Clark,
\textit{NIST Handbook of Mathematical Functions} (Cambridge University Press, Cambridge, England, 2010).
%
\bibitem{Hollands:2001nf}
S.~Hollands and R.~M.~Wald,
%``Local Wick polynomials and time ordered products of quantum fields in curved spacetime,''
Commun. Math. Phys. \textbf{223}, 289-326 (2001) %doi:10.1007/s002200100540
[arXiv:gr-qc/0103074 [gr-qc]].
%
\bibitem{Duff:1977ay}
M.~J.~Duff,
%``Observations on Conformal Anomalies,''
Nucl. Phys. B \textbf{125}, 334-348 (1977). %doi:10.1016/0550-3213(77)90410-2
\end{thebibliography}
\end{document}